\documentclass[12pt,authoryear]{elsarticle}

\usepackage{hyperref}
\usepackage{graphicx}
\usepackage{subcaption}
\usepackage{amssymb}
\usepackage{amsmath}
\usepackage{multirow}
\usepackage[utf8]{inputenc}
\usepackage[depythontex]{pythontex}
\usepackage{pgfplots}
\usepackage{pgfplotstable}
\usepgfplotslibrary{groupplots}
\usepackage{float} 
\usepackage{derivative}
\usepackage{latexcolors} 
\usepackage{import} 
\usepackage{siunitx}
\usepackage{catchfile}
\usepackage{mathrsfs} 
\usepackage{standalone} 
\usepackage{makecell}
\usepackage{algorithm}
\usepackage{algpseudocodex}
\usepackage{cleveref}
\usepackage{nicefrac}
\usepackage{threeparttable} 
\usepackage{booktabs}

\pgfplotsset{compat=newest,
	table/search path={figs} 
}

\usepgfplotslibrary{external}
\usetikzlibrary{math, shapes.geometric, shapes.misc, calc, arrows.meta, fit}
\usepackage{tcolorbox}

\tikzsetexternalprefix{figcache/}
\tikzset{external/mode=list and make}

\usepackage[abbreviations]{glossaries-extra}
\setabbreviationstyle{long-short} 

\usepackage{xcolor}
\usepackage{xargs}
\usepackage[colorinlistoftodos,textsize=footnotesize]{todonotes}
\usepackage{lineno}

\newcommandx{\unsure}[2][1=]{\todo[linecolor=red,backgroundcolor=red!25,bordercolor=red,#1]{#2}}
\newcommandx{\change}[2][1=]{\todo[linecolor=blue,backgroundcolor=blue!25,bordercolor=blue,#1]{#2}}
\newcommandx{\info}[2][1=]{\todo[linecolor=OliveGreen,backgroundcolor=OliveGreen!25,bordercolor=OliveGreen,#1]{#2}}

\usepackage{array}
\newcommand{\PreserveBackslash}[1]{\let\temp=\\#1\let\\=\temp}
\newcolumntype{C}[1]{>{\PreserveBackslash\centering}p{#1}}
\newcolumntype{R}[1]{>{\PreserveBackslash\raggedleft}p{#1}}
\newcolumntype{L}[1]{>{\PreserveBackslash\raggedright}p{#1}}

\newabbreviation
{sph}
{SPH}
{Smoothed Particle Hydrodynamics}

\newabbreviation
{wcsph}
{WCSPH}
{Weakly Compressible \gls{sph}}

\newabbreviation
{isph}
{ISPH}
{Incompressible \gls{sph}}

\newabbreviation
{mms}
{MMS}
{Method of Manufactured Solutions}

\newabbreviation
{sisph}
{SISPH}
{Simple Iterative \gls{sph}}

\newabbreviation
{dtsph}
{DTSPH}
{Dual-Time \gls{sph}}

\newabbreviation
{edac}
{EDAC}
{Entropically Damped Artificial Compressibility}

\newabbreviation
{dbc}
{DBC}
{Dynamic Boundary Condition}

\newabbreviation
{mdbc}
{mDBC}
{modified \gls{dbc}}

\newabbreviation
{lust}
{LUST}
{Local Uniform STencil}

\newabbreviation
{tvdrk2}
{TVD-RK2}
{Total Variaton Diminishing - Runge Kutta 2}

\newabbreviation
{fvm}
{FVM}
{Finite Volume Method}

\newabbreviation
{fou}
{FOU}
{First Order Upwind}

\newabbreviation
{mls}
{MLS}
{Moving Least Squares}

\newabbreviation
{tvf}%
{TVF}%
{Transport Velocity Formulation}

\newabbreviation
{epec}%
{EPEC}%
{Evaluate Predict Evaluate Correct}

\newabbreviation
{ccvt}%
{CCVT}%
{Capacity Constrained Voronoi Tessellations}
\journal{}

\begin{document}

\begin{frontmatter}

	\title{Rapid Variable Resolution Particle Initialization for Complex Geometries}
	\author[IITB]{Navaneet Villodi\corref{cor1}}
	\ead{navaneet@iitb.ac.in}
	\author[IITB]{Prabhu Ramachandran}
	\ead{prabhu@aero.iitb.ac.in}
	\address[IITB]{Department of Aerospace Engineering, Indian Institute of
		Technology Bombay, Powai, Mumbai 400076}

	\cortext[cor1]{Corresponding author}

	\begin{abstract}
		The accuracy of meshless methods like \gls{sph} is highly dependent on the quality of the particle distribution.
		Existing particle initialization techniques often struggle to simultaneously achieve adaptive resolution, handle intricate boundaries, and efficiently generate well-packed distributions inside and outside a boundary.
		This work presents a fast and robust particle initialization method that achieves these goals using standard \gls{sph} building blocks.
		Our approach enables simultaneous initialization of fluid and solid regions, supports arbitrary geometries, and achieves high-quality, quasi-uniform particle arrangements without complex procedures like surface bonding.
		Extensive results in both 2D and 3D demonstrate that the obtained particle distributions exhibit good boundary conformity, low spatial disorder, and minimal density variation, all with significantly reduced computational cost compared to existing approaches.
		This work paves the way for automated particle initialization to accurately model flow in and around bodies with meshless methods, particularly with \gls{sph}.
	\end{abstract}


\end{frontmatter}

\section{Introduction} \label{sec:intro}

Meshless methods require discretizing the domain into nodes.
Unlike traditional mesh-based methods, these nodes are scattered without fixed connectivity.
The nodes are required to be locally quasi-uniform and isotropic to ensure solution accuracy \citep{vandersandeFastVariableDensity2021,litvinovConsistenceConvergenceConservative2015,colagrossiParticlePackingAlgorithm2012}.
In the context of \gls{sph}, these nodes are referred to as particles.
Each particle is the representative of a (fuzzy) space surrounding itself.
The particles carry physical properties such as mass, position, velocity, and density.

Accurate mesh-free simulation of fluid flows involving boundaries require that the boundaries' intricacies be captured accurately in terms of distributed particles.
Most boundary treatment methods in \gls{sph} require that the particles be distributed on either side of the boundary as summarized by \citet{negiHowTrainYour2022a,villodiRobustSolidBoundary2024}.
Getting an initial particle distribution for this purpose with zero-order consistency at the interface between body and fluid is challenging.

Various authors have proposed methods to generate particle distributions.
\citet{perssonSimpleMeshGenerator2004} asservate that most mesh generation methods tend to be complex and inaccessible.
They propose a method in which an initial distribution of nodes is iteratively adjusted using Delaunay triangulations.
This method is simple, albeit with some numerics delegated to MATLAB's built-in functions.
\citet{colagrossiParticlePackingAlgorithm2012} proposed a novel particle packing algorithm, exclusively for initializing fluid particles.
They state that with an incompatible initialization, particles may tend to resettle to the static solution predicted by the SPH scheme in-simulation, thus emphasizing the importance of the algorithm being \gls{sph} native to avoid such spurious motions.

\citet{xiongGPUacceleratedAdaptiveParticle2013} proposed clipping a grid using the boundary.
The slit cells are assigned partial masses while the untouched cells are assigned full masses.
\citet{jiangBlueNoiseSampling2015} proposed a method to generate particle distributions with variable resolution.
They demonstrated their method in 3D, as well.
\citet{diehlGeneratingOptimalInitial2015} and \citet{velavelaALARICAlgorithmConstructing2018} also proposed methods to generate particle distributions with variable resolution in 3D.
These methods, however, do not generate packed particle distributions for a solid and the surrounding fluid simultaneously.

\citet{fornbergFastGeneration2D2015a} proposed an advancing front method for generating 2D particle distributions.
This non-iterative method supported variable density and irregular boundaries.
\citet{vandersandeFastVariableDensity2021} built upon this work, extending it to 3D.
This newer method utilizes a background grid and introduces corrections to generate higher quality particle distributions.

\citet{fuOptimalParticleSetup2019a} proposed a method with particle relaxation based on the gradient of the pressure term of the momentum equation.
The pressure is derived from an equation of state they proposed.
Temporally reconstructed ghost particles based on level-set are used for dealing with boundaries.
\citet{fuIsotropicUnstructuredMesh2019} also proposed to constrain the motion of particles at the boundary surface and to project the near-surface particles to the boundary surface. \citet{zhuCADcompatibleBodyfittedParticle2021} were able to demonstrate packing on various geometries with simple surface bonding, i.e., projecting near-surface particles to the boundary surface without restricting their motion or using ghost particles. \citet{jiFeatureawareSPHIsotropic2021} introduced a correction term to account for the support at the boundary in lieu of the ghost particles. \citet{yuLevelsetBasedPreprocessing2023a} introduced yet another correction term based on a smoothed Heaviside function to account for the support at the boundary.
Recently, \citet{zhao2025physicsdriven} extended this lineage of work to support multi-body systems and simultaneous initialization of fluid and solid particles.

Another method that supports simultaneous initialization of fluid and solid particles was proposed by \citet{negiAlgorithmsUniformParticle2021}.
The hybrid method they proposed synthesizes the approaches of \citet{colagrossiParticlePackingAlgorithm2012} and \citet{jiangBlueNoiseSampling2015} with their improvements.
This method also involves the projection of the near-surface particles to the boundary surface and constraining the motion of particles at the boundary surface.
However, we find that their force based on Lennard-Jones like potential is a deterrent for introducing variable resolution support.

To summarize, the following are the common pain points that one encounters when dealing with the present methods:
\begin{enumerate}
	\item Simultaneous initialization: Separate initialization may lead to the formation of undesirable artefacts like gaps, especially around sharp features~\citep{zhao2025physicsdriven}.
	      Most existing methods do not support simultaneous initialization of fluid and solid particles, except some newer methods such as that of \citet{negiAlgorithmsUniformParticle2021, zhao2025physicsdriven}.
	\item Adaptive resolution: Support for adaptive resolution is crucial for efficiently resolving intricate features.
	      Some older methods support adaptive resolution, but the newer methods leave it to be desired.
	      For example, the method \citet{negiAlgorithmsUniformParticle2021} has no support for adaptive resolution and support for adaptive resolution in \citet{zhao2025physicsdriven} appears to be rudimentary.
	\item Complexity: Static confinement~\citep{yuLevelsetBasedPreprocessing2023a}, surface bonding~\citep{zhuCADcompatibleBodyfittedParticle2021, negiHowTrainYour2022a,zhao2025physicsdriven}, special treatment of sharp corners~\citep{negiAlgorithmsUniformParticle2021}, restricted movement of particles at the boundary~\citep{fuIsotropicUnstructuredMesh2019,negiHowTrainYour2022a}, alternate relaxation of fluid and solid particles~\citep{zhao2025physicsdriven} are essential to the existing methods.
	      These procedures are complex, and the implementation is difficult.
	\item Efficiency and Speed: Advancing front methods~\citep{fornbergFastGeneration2D2015a, vandersandeFastVariableDensity2021} have limited scope for parallelization.
	      Moreover, stemming from the previous point, complexity often brings performance overhead.
	      Therefore, relaxation-based methods that employ complex procedures are expected to be slow.
\end{enumerate}

We propose a method for particle initialization in SPH that addresses all the above pain points.
It is very efficient, allowing one to generate variable resolution particle initializations simultaneously for solids and surrounding fluids with comparable accuracy to existing methods.
Moreover, the method consists of pure \gls{sph} building blocks.
Therefore, it should be relatively easy for an \gls{sph} practitioner to implement within any SPH framework (or standalone).
In the following section, we describe the method in detail so that the reader can easily understand and implement it.

\section{Methodology} \label{sec:method}

We employ a setup similar to \citet{negiAlgorithmsUniformParticle2021} with three sets of particles, free particles, interface particles, and frozen particles, as shown in \cref{fig:particle-sets}.
The interface particles are the particles at the interface and carry outward normals to the interface.
The frozen particles form the boundary, and complete the kernel support for the free particles.
The interface and the frozen particles do not move.

\IfFileExists{figs_cache_particle-sets.pdf}{
	\begin{figure}
		\centering
		\includegraphics[width=0.45\textwidth]{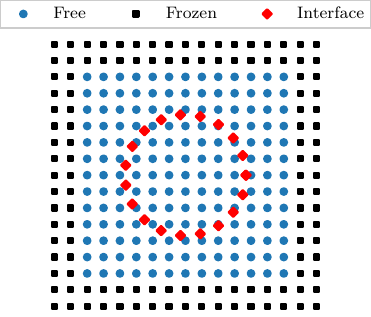}
		\caption{Particle sets used in the method.
			The blue circles represent free particles, the red diamonds represent interface particles, and the black squares represent frozen particles.
		}
		\label{fig:particle-sets}
	\end{figure}
}{\textcolor{red}{figure particle-sets.pdf not found!}}

Our proposed method consists of five components: restoring force, particle shifting, volume adaptivity, mass dissipation, and interface handling.
These components, along with other implementation details, are explained in this section.
\subsection{Restoring force} \label{sec:restoring-force}
We assume that the geometry to be packed is a shell in a pressurized fluid container.
Thus, we can make use of the familiar equation of motion for fluids,
\begin{equation}
	\label{eq:mom}
	\odv{\boldsymbol{u}}{t} = -\frac{1}{\rho}\nabla p
\end{equation}
where $\boldsymbol{u}$ is the velocity, $\rho$ is the density, and $p$ is the pressure. The pressure and density are related as
\begin{equation}
	\label{eq:eos}
	p = p_o \left(\frac{\rho}{\rho_o}\right)^{\gamma},
\end{equation}
where $p_o$ is the reference pressure and $\rho_o$ is the reference density. $p_o$, $\rho_o$, and $\gamma$ are constants.
The results in this paper are obtained with $p_o = 1$ Pa and $\rho_o = 1$ kg/m$^3$.
A higher $p$ would result in smaller time steps, $\Delta t$.
However, this has a negligible effect on the number of iterations required for convergence.
The choice of $\gamma$, on the other hand, is critical.
On a perfectly regular lattice arrangement, both density and kernel gradient sum~(ref.
eq. \ref{eq:kernel-gradient-sum}) approach their ideal values.
However, perfect regularity is not achievable when particles are to be packed around a curved interface.
In this scenario, the role of $\gamma$ is to prioritize one over the other.
Effectively, a higher $\gamma$ implies a stiffer equation of state.
For example, a $\gamma \approx 10$ emulates incompressible behaviour, minimizing the density variations.
On the other hand, a $\gamma \approx 1$ implies a softer equation of state, allowing for larger density variations.
Thus, a higher $\gamma$ prioritizes minimization of density errors over the kernel gradient sum.

One could also use a different equation in place of \cref{eq:eos}.
An interesting substitute is the one from the work of \citet{jiangBlueNoiseSampling2015}.
Their equation of state allows adjustment of $\rho_o$ to weigh in on a target profile where $\rho_i\to \rho_0$ and a target profile where $\sum_j \nabla W_{ij} m_j / \rho_j \to 0$.

\Cref{eq:mom} is discretized in \gls{sph} parlance as
\begin{equation}
	\label{eq:mom-sph}
	\odv{\boldsymbol{u}_i}{t} = -\frac{1}{\rho_i}\sum_{j \in \mathcal{N}_i} \left(p_i + p_j\right) \nabla W_{ij} \frac{m_j}{\rho_j},
\end{equation}
where the subscripts $i$ and $j$ are the particle indices. $j \in \mathcal{N}_i$, where $\mathcal{N}_i$ is the set of neighbours of particle $i$.
In further text, we use only $j$ for brevity.
$W_{ij}$ is the shorthand for the SPH kernel $W(\boldsymbol{r}_{ij}, h_{ij})$, where $h$ is the smoothing length, $h_{ij} = \left(h_i + h_j\right)/2$, $\boldsymbol{r}$ represents the position vector and $\boldsymbol{r}_{ij} = \boldsymbol{r}_i - \boldsymbol{r}_j$.
Similarly, $W(\boldsymbol{r}_{ij}, h_{i})$ would be represented by $W_i$.
\citet{sunMultiresolutionDeltaSPHTensile2018} has shown that the usage of $\left(p_i + p_j\right)$ in pressure gradient discretization accords regularizing inter-particle forces, encouraging particle settlement towards uniform spatial configurations.

\Cref{eq:eos} is straightforwardly discretized as
\begin{equation}
	\label{eq:eos-sph}
	p_i = p_o \left(\frac{\rho_i}{\rho_o}\right)^{\gamma}.
\end{equation}
The density of the particle, $\rho_i$, is obtained using summation density with iterative solution for smoothing lengths.
Equations
\begin{equation}
	\rho_i = \sum_{j} m_j W_{ij},
\end{equation}
and
\begin{equation}
	h_i = h_{\text{fact}} \left(\frac{1}{\sum_j W_{i}}\right)^{1/d}
\end{equation}
are solved iteratively using the Newton-Raphson method~\citep{priceSmoothedParticleHydrodynamics2012,hopkinsGeneralClassLagrangian2013,puriComparisonSPHSchemes2014}.
Here, $d$ is the number of dimensions and $h_{\text{fact}}$ is the ratio of smoothing length to particle spacing, $h/\sqrt[d]{m/\rho}$.

\subsection{Particle Shifting}
A particle shifting technique commonly used in \gls{sph} is used to regularize the distribution of particles~\citep{lindIncompressibleSmoothedParticle2012}.
We are using particle shifting in conjunction with the restoring force from \cref{sec:restoring-force} to make our algorithm faster.

The particle displacement, \(\delta \mathbf{r}_i\) is given as
\begin{equation}
	\delta \mathbf{r}_i =
	\begin{cases}
		-0.5 h^2 \widehat{\nabla} C_i                                                        & \text{if } 0.5 h^2 \left|\left|\widehat{\nabla} C_i\right|\right|_2 < 0.2 h \\
		-0.2 h \frac{\widehat{\nabla} C_i}{\left|\left|\widehat{\nabla} C_i\right|\right|_2} & \text{otherwise}
	\end{cases}
	,
\end{equation}
where
\begin{equation}
	\label{eq:grad-conc}
	\widehat{\nabla}
	C_i = \sum_{j}\left[1 + 0.2 \left(\frac{W_{ij}}{W(\Delta x_i)}\right)^4\right] \frac{m_j}{\rho_0} \nabla W_{ij}.
\end{equation}
Here, \(\Delta x\) is taken as the smoothing length times the inflection point of the kernel function.
Note the use of the reference density, \(\rho_0\), in the denominator in \cref{eq:grad-conc}, in place of the usual, \(\rho_i\).
To our knowledge, this modification was coined by \citet{mutaEfficientAccurateAdaptive2022} for some of their initialization requirements.
This has not been communicated in text but can be verified from their code repository available at \url{https://gitlab.com/pypr/adaptive_sph}.
In the present method, the restoring force works towards ironing out the local density variations.
With the usage of reference density in \cref{eq:grad-conc}, the particle shifting works towards zeroth-order consistency discounting the local density variations.
This helps us accelerate the convergence of the particle initialization method.

\subsection{Volume Adaptivity} \label{sec:volume-adaptivity}
Volume adaptivity allows us to generate variable resolution particle distributions.
This is achieved by splitting and merging particles.
For this, we borrowed from the algorithm of \citet{sunAccurateSPHVolume2021} and implemented it within the framework of \citet{haftuParallelAdaptiveWeaklycompressible2022}.
The adaptivity procedure is scheme agnostic and its derivates have been used with various SPH schemes~\citep{villodiSmoothedParticleHydrodynamics2025,villodiAdaptiveCompressibleSmoothed2025}.
The overall procedure is listed in \cref{alg:adapt-main}.
All the loops over free particles, and the procedures \textsc{UpdateVolumeBands}, \textsc{ProcessMerge}, and \textsc{InitOffspring} can be parallelized.
These procedures are expanded further in \cref{alg:update-volume-bands,alg:offspring-init,alg:find-merge-partner} and are also explained in further subsections, \labelcref{sec:splitting,sec:merging,sec:spatial-solution-adaptivity}.

\begin{algorithm}
	\caption{The main adaptive refinement procedure}%
	\label{alg:adapt-main}
	\begin{algorithmic}[1]
		\Procedure{AdaptMain}{}
		\ForAll{\(i \in\) free particles} \label{alg:adapt-main-outer-loop1}
		\ForAll{\(j \in\) interface particles neighbouring \(i\)}
		\State{$\Delta s_i = \min(\Delta s_{\min, j}, \Delta s_i)$}         \Comment{Inherit Spacing}
		\EndFor{}
		\EndFor{}
		\State{\Call {UpdateVolumeBands()}{}} \label{alg:adapt-main-update-bands}
		\ForAll{\(i \in\) free particles} \label{alg:adapt-main-outer-loop2}
		\If {\(i\) is merge worthy}
		\State{\Call {FindMergePartner()}{}} \label{alg:adapt-main-find-merge1}
		\EndIf{}
		\EndFor{}
		\State{\Call{ProcessMerge()}{}} \label{alg:adapt-main-process-merge1} \Comment{Process Merge Partners}
		\State{Add or remove particles}
		\State{\Call{InitOffspring()}{}} \label{alg:adapt-main-init-offspring} \Comment{Initialize Offspring Particles}
		\EndProcedure
	\end{algorithmic}
\end{algorithm}

\subsubsection{Splitting} \label{sec:splitting}

The splitting pattern is shown in \cref{fig:splitting-pattern}.
The \(n_o\) offspring particles are placed at the vertices of a regular $d$-dimensional hypercube of size \(a\) centred at the parent particle.
\begin{equation}
	a = \frac{\sqrt[d]{\frac{m_p}{\rho_p}}}{n_o},
\end{equation}
where \(n_o = 2^d\) is the number of offspring.
The subscripts $p$ and $o$ denote parent and offspring, respectively.
\begin{figure}
	\centering
	\begin{subfigure}[b]{0.3\textwidth}
		\centering
\begin{tikzpicture}[scale=2]
	\coordinate (O) at (0, 0);
	\coordinate (A) at (-0.5, -0.5);
	\coordinate (B) at (-0.5, 0.5);
	\coordinate (C) at (0.5, 0.5);
	\coordinate (D) at (0.5, -0.5);

	\fill[gray] (O) circle (5pt);
	\draw[opacity=0.2] (A) -- (B) -- (C) -- (D) -- cycle;
	\foreach \point in {A, B, C, D}
		{
			\fill[teal, opacity=0.9] (\point) circle (2.5pt);
		}
\end{tikzpicture}
	\end{subfigure}
	\begin{subfigure}[b]{0.3\textwidth}
		\centering
\begin{tikzpicture}[scale=2]
	\coordinate (O) at (0, 0, 0);
	\coordinate (A) at (-0.5, 0.5, 0.5);
	\coordinate (B) at (-0.5, -0.5, 0.5);
	\coordinate (C) at (0.5, -0.5, 0.5);
	\coordinate (D) at (0.5, 0.5, 0.5);
	\coordinate (E) at (-0.5, 0.5, -0.5);
	\coordinate (F) at (-0.5, -0.5, -0.5);
	\coordinate (G) at (0.5, -0.5, -0.5);
	\coordinate (H) at (0.5, 0.5, -0.5);
	\draw[opacity=0.2] (A) -- (B) -- (C) -- (D) -- cycle;
	\draw[opacity=0.2] (A) -- (E);
	\draw[opacity=0.2] (B) -- (F);
	\draw[opacity=0.2] (C) -- (G);
	\draw[opacity=0.2] (D) -- (H);
	\draw[opacity=0.2] (E) -- (F) -- (G) -- (H) -- cycle;
	\foreach \point in {A, B, C, D}
		{
			\shade[ball color=teal, opacity=0.9] (\point) circle (2.5pt);
		}
	\shade[ball color=gray] (O) circle (5pt);
	\foreach \point in {E, F, G, H}
		{
			\shade[ball color=teal, opacity=0.9] (\point) circle (2.5pt);
		}
\end{tikzpicture}
	\end{subfigure}
	\caption{Splitting pattern in 2D~(left) and 3D~(right).
		The grey parent particle is split into four offspring particles in 2D and eight offspring particles in 3D.
		The offspring particles are teal colored.
	}
	\label{fig:splitting-pattern}
\end{figure}

The mass and volume of the parent particle are distributed equally among the offspring particles.
Other properties like density, pressure and velocity are copied from the parent particle to the offspring particles.

\begin{algorithm}
	\caption{Initalise Offspring Particles}%
	\label{alg:offspring-init}
	\begin{algorithmic}[1]
		\Procedure{InitOffspring}{}
		\ForAll{\(p \in\) split worthy fluid particles}
		\State{\(a \leftarrow \sqrt[d]{m_p / \rho_p}/ n_o\)}
		\For{$k = 0, \dots, n_o - 1$}
		\LComment{Place the k\textsuperscript{th} offspring particle}
		\State{\(\Delta \boldsymbol{r}_{o,k} \leftarrow [0, 0, 0]\)}
		\For{\(i = 0, \dots, d - 1\)}
		\If{\(\mathrm{floor}(k/2^i)\) mod 2 = 0}
		\State{\(\Delta \boldsymbol{r}_{o,k}[i] \leftarrow - a/2\)}
		\Else
		\State{\(\Delta \boldsymbol{r}_{o,k}[i] \leftarrow a/2\)}
		\EndIf
		\EndFor
		\State{\(\boldsymbol{r}_{o,k} \leftarrow \boldsymbol{r}_p + \Delta \boldsymbol{r}_{o,k}\)}
		\Statex{}
		\LComment{Assign properties}
		\State{\(V_{o,k} \leftarrow V_p / n_o\)}
		\State{\(m_{o,k} \leftarrow m_p / n_o\)}
		\State{\(h_{o,k} \leftarrow h_p / 2\)}
		\State{\(\rho_{o,k} \leftarrow \rho_p\)}
		\State{\(\odv{\boldsymbol{u}_{o,k}}{t} \leftarrow \odv{\boldsymbol{u}_p}{t}\)}
		\State{\(\boldsymbol{u}_{o,k} \leftarrow \boldsymbol{u}_p\)}
		\State{\(p_{o,k} \leftarrow p_p \)}
		\EndFor \EndFor \EndProcedure
	\end{algorithmic}
\end{algorithm}

\subsubsection{Merging} \label{sec:merging}
Merging is performed pairwise, i.e., two particles identified as each other's merge partners merge to form one.
This is simple and can be performed in parallel.

The process of finding merge partners is described in \cref{alg:find-merge-partner}.
Here, $\left|\left|\boldsymbol{r}_{ij}\right|\right|$ is the distance between particles \(i\) and \(j\).
One of the particles is assigned the properties of the merged particle, and the other is marked for deletion as shown in \cref{alg:process-merge}.
\begin{algorithm}
	\caption{Find merge partner for merge-worthy particles}
	\label{alg:find-merge-partner}
	\begin{algorithmic}[1]
		\Procedure{FindMergePartner}{}
		\State{\(r_{ij,\min} \leftarrow \inf\)}
		\ForAll{\(j \in\) merge worthy fluid particles in neighbourhood of \(i\)}
		\If{\(\left|\left|\boldsymbol{r}_{ij}\right|\right|_2 <  \min{\left(r_{ij,\min}, \sqrt[d]{V_{\min,i}}\right)}\) }
		\State{\(r_{ij,\min} \leftarrow \left|\left|\boldsymbol{r}_{ij}\right|\right|_2\)}
		\State{\(mp_i \leftarrow j\)} \Comment{mp stands for merge partner}
		\EndIf
		\EndFor
		\EndProcedure
	\end{algorithmic}
\end{algorithm}

\begin{algorithm}
	\caption{Process merge partners}
	\label{alg:process-merge}
	\begin{algorithmic}[1]
		\Procedure{ProcessMerge}{}
		\ForAll{\(i \in\) merge worthy fluid particles}
		\State{\(j \leftarrow mp_i\)}
		\If{\(mp_j =i\)}
		\If{\(i < j\)} \Comment{Merge into \(i\)}
		\State{\(m \leftarrow m_i + m_j\)}
		\State{\(\rho_i \leftarrow m / (m_i / \rho_i + m_j / \rho_j)\)}
		\State{\(m_1 \leftarrow m_i / m\)}
		\State{\(m_2 \leftarrow m_j / m\)}
		\State{\(m_i \leftarrow m\)}
		\State{\(h_i \leftarrow \sqrt[d]{h_i^d + h_j^d}\)}
		\State{\(\boldsymbol{r}_i \leftarrow m_1 \boldsymbol{r}_i + m_2 \boldsymbol{r}_j\)}
		\State{\(\boldsymbol{u}_i \leftarrow m_1 \boldsymbol{u}_i + m_2 \boldsymbol{u}_j\)}
		\State{\(\odv{\boldsymbol{u}_i}{t} \leftarrow m_1 \odv{\boldsymbol{u}_i}{t} + m_2 \odv{\boldsymbol{u}_j}{t}\)}
		\State{Find \(p_i\) using the equation of state}
		\Else
		\State{Mark \(i\) for deletion}
		\EndIf
		\EndIf
		\EndFor
		\EndProcedure
	\end{algorithmic}
\end{algorithm}

\subsubsection{Volume Thresholds} \label{sec:spatial-solution-adaptivity}

Particles are initialized with the maximum expected reference spacing.
We define a particle property, \(\Delta s_{\min}\).
Note that $\Delta s_{\min, i} \neq \min_i(\Delta s_i)$ and is a property carried by each interface particle.
In each iteration, the reference spacing of the free particles is updated based on the interface particles as,
\begin{equation}
	\Delta s_i = \min(\Delta s_{\min, j}, \Delta s_i),
\end{equation}
where $j$ represents the interface particles in the neighbourhood of $i$.

One needs to create refinement bands that ensure a smoother transition between the fine and coarse regions to reduce interpolation errors.
A procedure to automate this is given by \citet{yangAdaptiveResolutionMultiphase2019,mutaEfficientAccurateAdaptive2022,haftuParallelAdaptiveWeaklycompressible2022}.
This procedure uses the refinement ratio parameter, \(C_r\).
Refinement ratio is defined as the ratio of the reference spacing between adjacent bands.
Let \(\Delta s_k\) be the reference spacing of the \(k\)\textsuperscript{th} band and \(\Delta s_{k+1}\) the reference spacing of the adjacent coarser \(k+1\)\textsuperscript{th} band.
Then, the refinement ratio would be
\begin{equation}
	C_r = \frac{\Delta s_{k+1}}{\Delta s_k}.
\end{equation}
We observe that a refinement ratio of 1.2 works well for most cases.
Once the refinement ratio is set, \cref{alg:update-volume-bands} forms $\Delta s$ bands, avoiding an abrupt transition between the refined and unrefined regions.
The thresholds for split and merge are also set by \cref{alg:update-volume-bands} using this $\Delta s$.

\begin{algorithm}
	\caption{Update volume bands}
	\label{alg:update-volume-bands}
	\begin{algorithmic}[1]
		\Procedure{UpdateVolumeBands}{}
		\ForAll{\(i \in\) in fluid particles}
		\State{Find \(\min_j{\left(\Delta s\right)}\), minimum \(\Delta s\) within neighbours of \(i\)}
		\State{Find \(\max_j{\left(\Delta s\right)}\), maximum \(\Delta s\) within neighbours of \(i\)}
		\State{Find \(\text{mean}_j{\left(\Delta s\right)}\), geometric mean \(\Delta s\) within neighbours of \(i\)}
		\If {\(\left(\max_j{\left(\Delta s\right)} / \min_j{\left(\Delta s\right)}\right) > \left(C_r\right)^3 \)}
		\State{\(\Delta s_i\) \(\leftarrow \min{\left(\max_j{\left(\Delta s\right)}, C_r \min_j{\left(\Delta s\right)}\right)}\)}
		\Else
		\State{\(\Delta s_i\) \(\leftarrow \text{mean}_j\Delta s\)}
		\EndIf

		\State{\(V_{\max,i} \leftarrow {8 (\Delta s_i)}^d / 5\)}
		\State{\(V_{\min,i} \leftarrow  {2 (\Delta s_i)}^d / 3 \)}
		\EndFor
		\EndProcedure
	\end{algorithmic}
\end{algorithm}

\subsection{Mass Dissipation} \label{sec:mass-dissipation}
As seen in \cref{sec:spatial-solution-adaptivity}, \(\Delta s\) varies in steps of \(C_r\), not smoothly.
Moreover, the particle volume can vary between 8/5 and 2/3 of ${\Delta s}^d$ as shown in \cref{alg:update-volume-bands}.
While \cref{alg:update-volume-bands} ensures that the particles of wildly different sizes do not interact with each other, there can still be irregularities arising out the jumps in \(\Delta s\) across the spacing bands, or other issues with adaptive refinement, say an eligible particle not being able to find a merge partner.
These are effectively mitigated with the mass dissipation from \citet{prasannakumarMultimassCorrectionMulticomponent2018}.
In essence, the interacting particles with different masses exchange a small portion of their mass as
\begin{equation}
	\label{eq:mass-exchange}
	\frac{\mathrm{d} m_i}{\mathrm{d} t} = \sum_{j} \frac{m_j + m_i}{\rho_j + \rho_i} \psi_{ij} v_{ij}^{sig,m} m_{ij} \hat{\mathbf{r}}_{ij} \cdot \nabla_i W_{ij},
\end{equation}
where the signal velocity, \(v_{ij}^{sig,m}\) is given as
\begin{equation}
	v_{ij}^{sig,m} =
	\begin{cases}
		\sqrt{\frac{\left|p_{ij}\right|}{\rho_i + \rho_j}} & \text{if } \hat{\mathbf{r}}_{ij} \cdot \mathbf{v}_{ij} < 0 \\
		0                                                  & \text{otherwise}
	\end{cases}
\end{equation}
and the limiter function, \(\psi_{ij}\) is given as
\begin{equation}
	\psi_{ij} = \frac{\left(\mathbf{v}_{ij} \cdot \hat{\mathbf{r}}_{ij}\right)^2}{\mathbf{v}_{ij} \cdot \mathbf{v}_{ij} + \left(\epsilon_\psi v_{ij}^{sig,m}\right)^2 + \left(0.0005 (c_i + c_j)\right)^2},
\end{equation}
where \(\epsilon_\psi\) is set as \(0.01\). \(m_{ij} = m_i - m_j\) and similarly \(\mathbf{v}_{ij}\) and \(p_{ij}\). \(c = \sqrt{\gamma p / \rho}\), the speed of sound and \(\hat{\mathbf{r}}_{ij}\) is the unit vector in the direction of \(\mathbf{r}_{ij}\).
The quantities are symmetrized between particles \(i\) and \(j\) in the mass exchange formulation, ensuring global conservation of mass~\citep{prasannakumarMultimassCorrectionMulticomponent2018}. The mass exchange effectively smoothes out the local mass variations, which in turn smoothes out the local volume variations.

\subsection{Interface Handling} \label{sec:interface-handling}
Given a set of points representing the interface, particles near the interface are displaced away from the interface.
The displacement is just enough to ensure that there are no particles at a distance less than $\Delta s_d$, where
\begin{equation}
	\label{eq:interface-spacing}
	\Delta s_d =
	\begin{cases}
		\frac{\sqrt[4]{3}}{2\sqrt{2}} \Delta s & \text{in 2D} \\
		\frac{\sqrt[3]{4}}{2\sqrt{3}} \Delta s & \text{in 3D}
	\end{cases}
\end{equation}
The coefficients in \cref{eq:interface-spacing} are obtained using simple geometric arguments as explained in \ref{sec:appendix-repel}.

If the interface points are closely spaced, say with a spacing less than $0.1 \Delta s$, simply displacing free particles away from the nearest interface point would suffice.
Otherwise, an alternate strategy is required to displace the particles away from the interface.
Given the position vector of a particle, \(\boldsymbol{r}_i\), and the position vector of a neighbouring interface point, \(\boldsymbol{r}_j\), and the unit outward normal at the interface point, \(\hat{\boldsymbol{n}}_j\), the distance from the particle to the interface projected along $\hat{\boldsymbol{n}}_j$ would be $\boldsymbol{r}_{ij} \cdot \hat{\boldsymbol{n}}_j$.
If we imagine a separating margin of \(\Delta s_d\) to either side of the interface, then $\Delta s_d - \left|\boldsymbol{r}_{ij} \cdot \hat{\boldsymbol{n}}_j \right|$ if positive would give us the distance by which the particle is into the margin.
In other words, $\max{\left(\Delta s_d - \left|\boldsymbol{r}_{ij} \cdot \hat{\boldsymbol{n}}_j \right|, 0\right)}$ is the distance by which the particle needs to be displaced along or opposite to $\hat{\boldsymbol{n}}_j$ to be outside the margin.
With this, a displacement vector can be constructed by sampling the neighbouring interface points as
\begin{equation}
	\label{eq:interface-displacement}
	\delta \boldsymbol{r}_{d,i} =  \frac{ \sum_j \hat{\boldsymbol{n}}_j \max{\left(\Delta s_d -   \left|\boldsymbol{r}_{ij} \cdot \hat{\boldsymbol{n}}_j \right|, 0\right)} \frac{\boldsymbol{r}_{ij} \cdot \hat{\boldsymbol{n}}_j}{\left|\boldsymbol{r}_{ij} \cdot \hat{\boldsymbol{n}}_j\right|}
		W_{i}}{\sum_j W_{i}}.
\end{equation}
Here, $\boldsymbol{r}_{ij} \cdot \hat{\boldsymbol{n}}_j / \left|\boldsymbol{r}_{ij} \cdot \hat{\boldsymbol{n}}_j\right|$ dictates weather the particle should be moved along or opposite to $\hat{\boldsymbol{n}}_j$ and $W_i / \sum_j W_i$ acts like an inverse distance weighting function.

\subsection{Time integration and other details} \label{sec:time-integration}
Time integration is performed using the semi-implicit Euler method,
\begin{equation}
	\begin{aligned}
		\boldsymbol{u}_i\left(t + \Delta t\right) & = \boldsymbol{u}_i\left(t\right) + \Delta t \odv{\boldsymbol{u}_i}{t}\left(t\right) ,
		\\
		\boldsymbol{r}_i\left(t + \Delta t\right) & = \boldsymbol{r}_i\left(t\right) + \Delta t \boldsymbol{u}_i\left(t + \Delta t\right),
	\end{aligned}
\end{equation}
The time step is determined as
\begin{equation}
	\Delta t = C_{CFL} \min\left(\Delta t_\text{vel}, \Delta t_\text{force}\right).
\end{equation}
where
\begin{equation}
	\Delta t_\text{vel} = \frac{h_\text{min}}{\max(c)}
\end{equation}
and
\begin{equation}
	\Delta t_\text{force} = C_\text{force} \sqrt{\frac{h_\text{min}}{\max(\left|\left|\odv{\boldsymbol{u}}{t}\right|\right|_2)}},
\end{equation}

The cubic spline~\citep{monaghanSmoothedParticleHydrodynamics1992} kernel is used.
The smoothing length is set with $h_\text{fact} = 1.2$.
The algorithm is implemented in the \texttt{PySPH}~\citep{ramachandranPySPHPythonbasedFramework2021a} framework.
The simulations are orchestrated using \texttt{automan}~\citep{ramachandranAutomanPythonBasedAutomation2018}.
The source is available at \url{https://gitlab.com/pypr/particle-packing}.

\section{Results} \label{sec:results}

This section presents the results of the particle initialization method described in \cref{sec:method}.
Firstly, we show the results of relaxation in an initially perturbed set of particles without an interface.
\Cref{fig:stability-nobody} shows the initial configuration and final distribution of particles for the case of an initialization with no interface.
Similarily, \cref{fig:stability-hex} shows the initial and final particle distributions for the case of an initialization with a hexagonal interface.
The particle spacing is carefully chosen so that the hexagonal interface falls midway two layers as shown in \cref{fig:stability-hex}.
\Cref{fig:stability-nobody} and \cref{fig:stability-hex} show that the particles arrange themselves in a regular pattern if there is no interface or if the interface is aligned with the particle layers.

\IfFileExists{figs_cache_Stability_stability_nobody.pdf}{
	\begin{figure}
		\centering
		\includegraphics[width=0.95\textwidth]{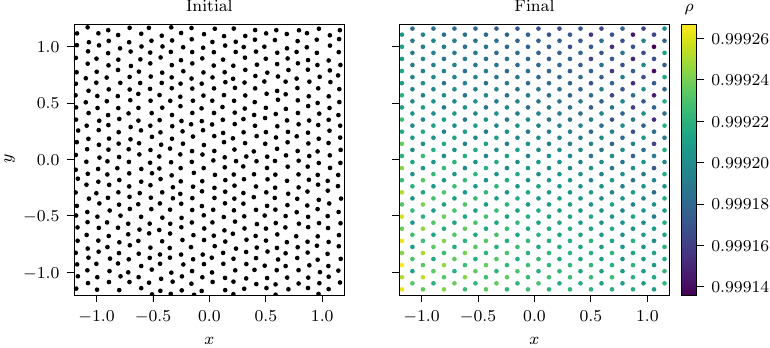}
		\caption{Particle initialization without interface, initial configuration is on the left and the final configuration is coloured by $\rho$ is on the right.}
		\label{fig:stability-nobody}
	\end{figure}
}{\textcolor{red}{Figure stability nobody not found}}

\IfFileExists{figs_cache_Stability_stability_hex.pdf}{
	\begin{figure}
		\centering
		\includegraphics[width=0.95\textwidth]{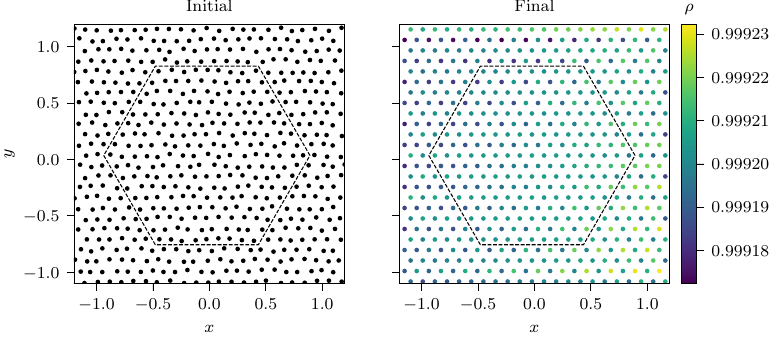}
		\caption{Particle initialization around a hexagon, initial configuration is on the left and the final configuration is coloured by $\rho$ is on the right.}
		\label{fig:stability-hex}
	\end{figure}
}{\textcolor{red}{Figure stability hex not found}}

We present the results of the particle initialization on variety of geometries as listed below:
\begin{itemize}
	\item {2D}
	      \begin{itemize}
		      \item Constant Resolution
		            \begin{itemize}
			            \item Circle
			            \item Starfish
		            \end{itemize}
		      \item Variable Resolution
		            \begin{itemize}
			            \item Fire Emoji
			            \item NACA 0012 Airfoil
		            \end{itemize}
	      \end{itemize}
	\item {3D}
	      \begin{itemize}
		      \item Constant Resolution
		            \begin{itemize}
			            \item Ellipsoid
			            \item Stanford Bunny
		            \end{itemize}
		      \item Variable Resolution
		            \begin{itemize}
			            \item Utah Teapot
			            \item Onera Wing
			            \item Ship
		            \end{itemize}
	      \end{itemize}
\end{itemize}
The 2D examples are run on an 6 core Intel i7-8700 desktop with 16 GB RAM and the 3D examples are run on 48 core dual socket Intel Xeon Gold 6240R compute cluster node with 192 GB RAM.
Following the examples, we present supersonic flow over a biconvex airfoil as a validation case.
Finally, we present a discussion on the performance of this method and the presented results.

We use the following parameters to quantify the quality of the particle initialization for the above listed test cases:
\begin{itemize}
	\item Density: The density of the particles is computed using the summation density.
	      The density should tend to the reference density, $\rho_o$, for all particles.
	      The uniformity of the density implies uniformity of pressure, by virtue of the equation of state, \cref{eq:eos}.
	      Uniformity of pressure, in turn implies balanced forces on the particles, by \cref{eq:mom}.
	\item Kernel gradient sum: The kernel gradient sum is computed as
	      \begin{equation}
		      \label{eq:kernel-gradient-sum}
		      \nabla \Gamma_i = \sum_{j} \nabla W_{i} \frac{m_j}{\rho_j}.
	      \end{equation}
	      The evaluation of force on a particle in \cref{eq:eos-sph} involves the use of the kernel gradient.
	      Therefore, in a relaxed configuration, the kernel gradient sum should tend to zero~\citep{zhao2025physicsdriven}. 
		  This is a necessary condition to approximate a given function with $\mathcal{O}(h^2)$ accuracy.
	      We present the norm of the kernel gradient sum, \(\left|\left|\nabla \Gamma_i\right|\right|_2\) for the cases presented in this section.
	\item Spatial disorder measure: The spatial disorder measure was proposed by \citet{antuonoMeasureSpatialDisorder2014} to quantify how far a given particle distribution is from a regular arrangement.
	      The spatial disorder measure is computed using two distances.
	      The first distance is the distance to the nearest neighbour,
	      \begin{equation}
		      d_{1i} = \min_j (\left|\left|\boldsymbol{r}_{ij}\right|\right|_2).
	      \end{equation}
	      To obtain the second distance, 8 cones of half-angle \(\theta_c\) are considered around the particle,
	      \begin{equation}
		      \mathcal{C}_{i,k} = \left\{\boldsymbol{r}_l \in \mathbb{R}^d : \frac{\boldsymbol{r}_{li}}{\left|\left|\boldsymbol{r}_{li}\right|\right|_2} \cdot \hat{\boldsymbol{v}}_k \leq \cos(\theta_c)\right\},
	      \end{equation}
	      where $\theta_c = 7 \pi / 18$ and \(\hat{\boldsymbol{v}}_k\) is the unit vector representing the direction of the \(k\)\textsuperscript{th} cone.
	      \begin{equation}
		      \hat{\boldsymbol{v}} \in
		      \begin{cases}
			      \left\{ [\pm\sqrt{2}, \pm\sqrt{2}]\right\} \cup \left\{[\pm1, 0] \right\} \cup \left\{ [0, \pm1] \right\}
			       & \text{in 2D} \\
			      \left\{ [\pm\sqrt{2}, \pm\sqrt{2}, \pm\sqrt{2}] \right\}
			       & \text{in 3D}
		      \end{cases}
		      .
	      \end{equation}
	      The second distance is
	      \begin{equation}
		      d_{2i} = \max_{k} d_{2i,k}
	      \end{equation}
	      where $d_{2i,k}$ is the distance to the nearest neighbour within the cone \(\mathcal{C}_{i,k}\)
	      \begin{equation}
		      d_{2i,k} = \min_{\boldsymbol{r}_l \in \mathcal{C}_{i,k}} \left|\left|\boldsymbol{r}_{li}\right|\right|_2.
	      \end{equation}
	      The local spatial disorder measure is then defined as
	      \begin{equation}
		      \lambda_i =
		      \begin{cases}
			      \frac{d_{2i} - d_{1i}}{d_{1i} + d_{2i}} & \text{if } d_{1i} > 0 \\
			      0                                       & \text{otherwise}
		      \end{cases}
		      ,
	      \end{equation}
	      and the global spatial disorder follows as a summation over all particles as
	      \begin{equation}
		      \Lambda = \frac{\sum_i \lambda_i}{N},
	      \end{equation}
	      where \(N\) is the total number of particles.
\end{itemize}

While we use a cubic spline kernel with $h_{\text{fact}}=1.2$ for relaxation, we often use the quintic spline kernel~\citep{negiTechniquesSecondorderConvergent2022,mutaEfficientAccurateAdaptive2022,villodiRobustSolidBoundary2024,villodiAdaptiveCompressibleSmoothed2025} and \(h_{\text{fact}} = 1.5\)~\citep{puriComparisonSPHSchemes2014,sunAccurateSPHVolume2021, villodiRobustSolidBoundary2024,villodiAdaptiveCompressibleSmoothed2025} for \gls{sph} simulations.
Resampling the obtained particle distributions with the quintic spline kernel and \(h_{\text{fact}} = 1.5\) results in significantly lower errors.
We have also tabulated these for the cases presented in this section.

\subsection{Circle}
In this subsection, we present the results of the particle initialization of a circle with constant resolution in 2D.
A circle of unit radius is discretized with a constant resolution of \(\left(m/\rho\right)^{1/d}= 0.1\).
From \cref{fig:circle-disorder}, we observe that the particle initialization converges to a relaxed configuration with a spatial disorder measure, \(\Lambda\), of around 0.0225.
The density is comparable to the results of \citet{negiAlgorithmsUniformParticle2021}.
The caveat is that the particles are not as boundary-hugging as a surface bonding method would result in.
The particles, however, appear more regular in comparison, especially away from the interface.

\IfFileExists{figs_cache_Circle_noadapt.pdf}{
	\begin{figure}
		\centering
		\includegraphics[width=0.95\textwidth]{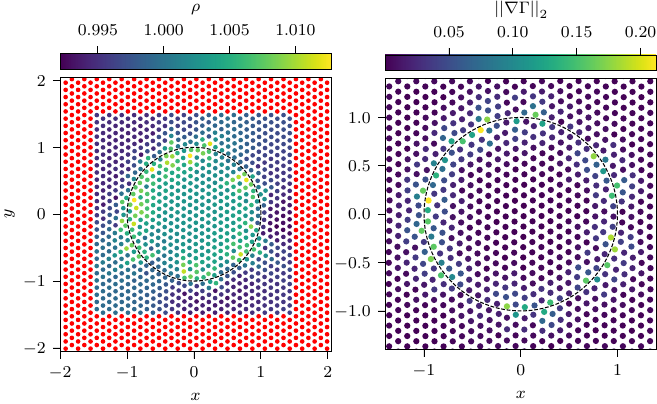}
		\caption{Particle initialization of a circle with constant resolution, coloured by $\rho$~(left) and $\left|\left|\nabla \Gamma\right|\right|_2$~(right).
			The plot on the left shows the complete domain with the frozen particles in red.
		}
		\label{fig:circle}
	\end{figure}
}{\textcolor{red}{Figure kgs not found}}

\IfFileExists{figs_cache_Circle_disorder.pdf}{
	\begin{figure}
		\centering
		\includegraphics[width=0.55\textwidth]{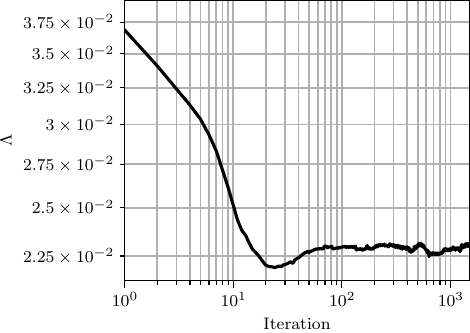}
		\caption{Spatial disorder measure, $\Lambda$, for the particle initialization of a circle with constant resolution.}
		\label{fig:circle-disorder}
	\end{figure}
}{\textcolor{red}{Figure disorder not found}}

\begin{figure}
	\centering
	\includegraphics[width=0.95\textwidth]{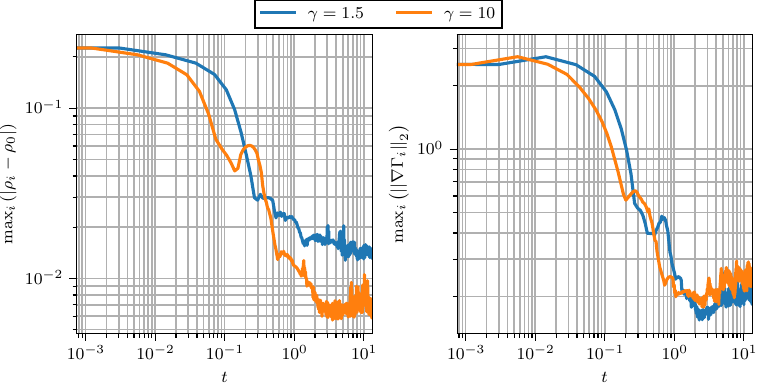}
	\caption{Convergence of the particle initialization of a circle with constant resolution, showing the maximum density error (left) and maximum kernel gradient sum (right) for different values of $\gamma$.}
	\label{fig:circle-conv}
\end{figure}

The error values are presented with the original cubic spline kernel and $h_{\text{fact}} = 1.2$ and the resampled quintic spline kernel and $h_{\text{fact}} = 1.5$ are presented in \cref{tab:circle-stats}.
As discussed in \cref{sec:method}, a higher $\gamma$ prioritizes minimization of $\rho$ errors over $\left|\left|\nabla \Gamma\right|\right|_2$.
This can be seen in \cref{fig:circle-conv}, where the maximum density error is lower for $\gamma = 10$ than for $\gamma = 1.5$, while the maximum kernel gradient sum is lower for $\gamma = 1.5$ than for $\gamma = 10$.
The same can also be seen in the error values presented in \cref{tab:circle-stats}.

\begin{table}
	\centering
	\begin{tabular}{cccccc}
	\toprule
	                        &                & $h_\text{fact}$ & $\max_i{\left(\left|\rho_i - \rho_0\right|\right)}$ & $\max_i{\left(\left|\left|\nabla \Gamma_i\right|\right|_2\right)}$ \\
	Gamma                   & Kernel         &                 &                                                     &                                                                    \\
	\midrule
	\multirow[t]{2}{*}{1.5} & Cubic Spline   & 1.2             & 0.0127                                              & 0.2123                                                             \\
	                        & Quintic Spline & 1.5             & 0.0084                                              & 0.0221                                                             \\
	\cline{1-5}
	\multirow[t]{2}{*}{10}  & Cubic Spline   & 1.2             & 0.0063                                              & 0.2393                                                             \\
	                        & Quintic Spline & 1.5             & 0.0027                                              & 0.0135                                                             \\
	\cline{1-5}
	\bottomrule
\end{tabular}

	\caption{Particle initialization error values for a circle with constant resolution.}
	\label{tab:circle-stats}
\end{table}

\subsection{Starfish}
The Starfish is a test case from \citet{negiAlgorithmsUniformParticle2021}, representing arbitrarily shaped geometries.
Here, we present the results of the particle initialization with a constant resolution of \(\Delta s = 0.1\).
The initialization procedure converges to a relaxed configuration with a spatial disorder measure, \(\Lambda\), of around 0.02 as shown in \cref{fig:starfish-disorder}.
The density variation and the kernel gradient sum are shown in \cref{fig:starfish}.
The density variation is comparable to the results of \citet{negiAlgorithmsUniformParticle2021}.
The convergence with varying spatial resolution is presented in \cref{fig:starfish-conv-sp}.
For this assessment, initialization was performed with \(\Delta s =\) 0.2, 0.1 and 0.05 till $t=10$.
The plotted quantity \(\mathbb{E}\left(\rho\right)\) is computed as
\begin{equation}
	\mathbb{E}\left(\rho\right) = \sqrt{\sum_i \left(\rho_i - \rho_0\right)^2 \hat{m}_i },
\end{equation}
where \(\hat{m}_i = m_i / \sum_i m_i\).
Both \(\Lambda\) and \(\mathbb{E}\left(\rho\right)\) exhibit approximately linear convergence with respect to the particle spacing.
The error values are presented with the original cubic spline kernel and $h_{\text{fact}} = 1.2$, and the resampled quintic spline kernel and $h_{\text{fact}} = 1.5$ are presented in \cref{tab:starfish-stats}.
With this, we conclude that the particle initialization method is suitable for arbitrarily shaped 2D geometries with constant resolution.

\IfFileExists{figs_cache_Starfish_noadapt.pdf}{
	\begin{figure}
		\centering
		\includegraphics[width=0.95\textwidth]{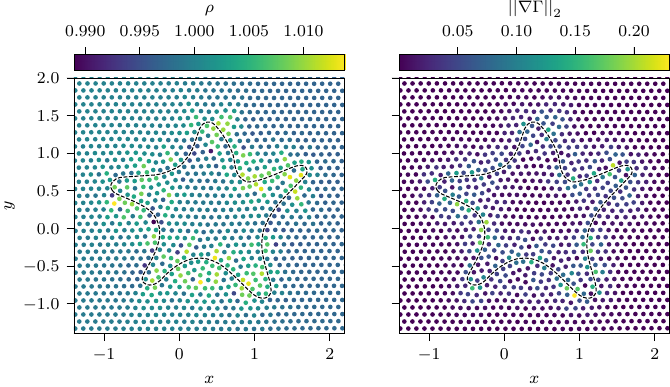}
		\caption{Particle initialization of a starfish with constant resolution, coloured by $\rho$~(left) and $\left|\left|\nabla \Gamma\right|\right|_2$~(right).}
		\label{fig:starfish}
	\end{figure}
}{\textcolor{red}{Figure noadapt not found}}

\IfFileExists{figs_cache_Starfish_disorder.pdf}{
	\begin{figure}
		\centering
		\includegraphics[width=0.6\textwidth]{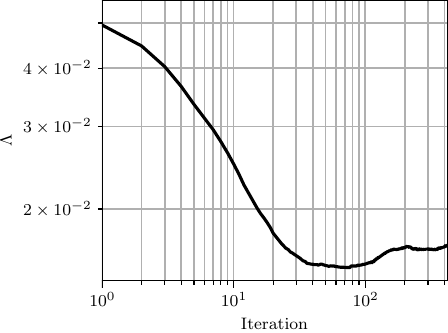}
		\caption{Spatial disorder measure, $\Lambda$, for the particle initialization of a starfish with constant resolution.}
		\label{fig:starfish-disorder}
	\end{figure}
}{\textcolor{red}{Figure disorder not found}}

\begin{table}
	\centering
	\begin{tabular}{ccccc}
	\toprule
	Kernel         & $h_\text{fact}$ & $\max_i{\left(\left|\rho_i - \rho_0\right|\right)}$ & $\max_i{\left(\left|\left|\nabla \Gamma_i\right|\right|_2\right)}$ \\
	\midrule
	Cubic Spline   & 1.2             & 0.0138                                              & 0.2306                                                             \\
	Quintic Spline & 1.5             & 0.0072                                              & 0.0407                                                             \\
	\bottomrule
\end{tabular}

	\caption{Particle initialization error values for a starfish with constant resolution.}
	\label{tab:starfish-stats}
\end{table}

\IfFileExists{figs_cache_Starfish_convsp.pdf}{
	\begin{figure}
		\centering
		\includegraphics[width=0.95\textwidth]{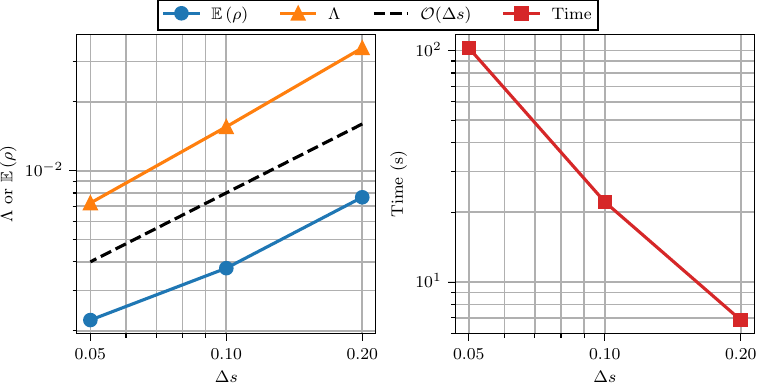}
		\caption{Convergence of the particle initialization of a starfish with constant resolution, showing $\Lambda$ and $\mathbb{E}\left(\rho\right)$ on the left and time taken for initialization on the right for different values of $\Delta s$.}
		\label{fig:starfish-conv-sp}
	\end{figure}
}{\textcolor{red}{Figure convsp not found}}

\subsection{Fire Emoji}

In this subsection, we present the results of the particle initialization of a fire emoji shape with variable resolution in 2D.
The domain is initially filled with particles at a constant resolution of \(\Delta s = 0.1\).

\Cref{fig:fire} demonstrates the effect of mass dissipation on the resulting particle distribution with the results with mass dissipation shown in the left column and without mass dissipation shown in the right column.
The top row of \cref{fig:fire} shows the particle distribution colored by density.
The frozen particles are also shown in red to demonstrate the extent of the domain and the arrangement of the free particles near the boundary.
The bottom row of \cref{fig:fire} shows the particle distribution colored by the kernel gradient sum.
At the points represented by red stars in plots in the bottom row of \cref{fig:fire}, $\Delta s_\text{min}$ is set to \(0.25 \Delta s\).
Comparing the two columns, we can see that the particle size varies smoothly in the case with mass dissipation, unlike the case without mass dissipation.
In the case without mass dissipation, at the edge of a spacing band, a density overshoot on the coarser side and an undershoot on the finer side can be observed.
The noisier distribution of particles in the case without mass dissipation results in higher kernel gradient sum errors as well.

The convergence of the spatial disorder measure, \(\Lambda\), with time is shown in \cref{fig:fire-disorder}.
The convergence with varying spatial resolution is presented in \cref{fig:fire-conv-sp}.
For this assessment, initialization was performed with \(\Delta s =\) 0.2, 0.1 and 0.05 till $t=10$.
The error values with and without mass dissipation are presented with the original cubic spline kernel with $h_{\text{fact}} = 1.2$, and the resampled quintic spline kernel with $h_{\text{fact}} = 1.5$ in \cref{tab:fire-stats}.
The error values with mass dissipation are significantly lower than those without mass dissipation.
Increased density variation and kernel gradient sum metrics are observed with variable resolution compared to constant resolution cases.
Nevertheless, it can also be clearly observed that the particle initialization method can create a particle distribution that is refined around the points of interest with smoothly varying volumes.
This is of practical importance in various applications.

\IfFileExists{figs_cache_Fire_adapt.pdf}{
	\begin{figure}
		\centering
		\includegraphics[width=0.95\textwidth]{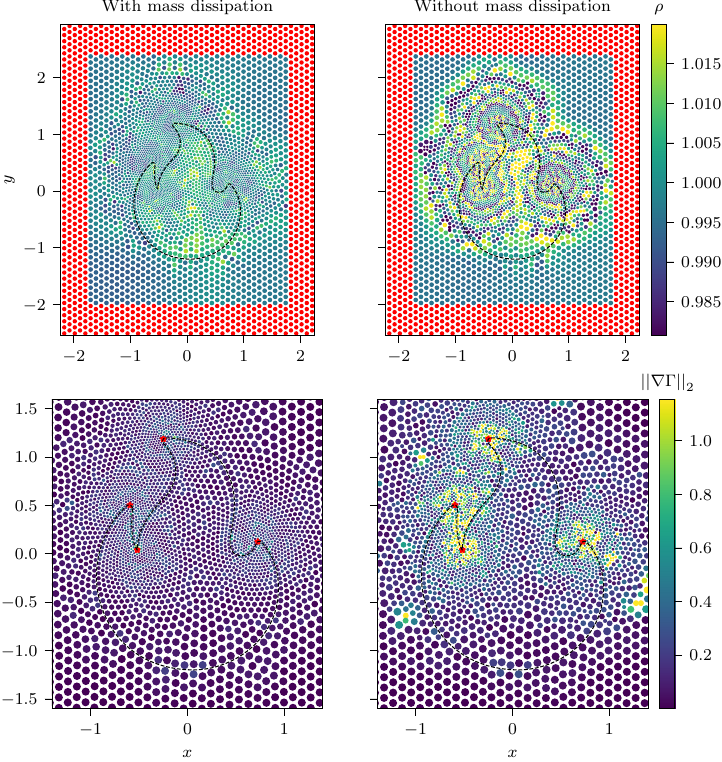}
		\caption{Particle initialization of a fire emoji with variable resolution, coloured by $\rho$~(top) and $\left|\left|\nabla \Gamma\right|\right|_2$~(bottom).
			The left column shows the results with mass dissipation, while the right column shows the results without mass dissipation.
			The plots on the top show the complete domain with the frozen particles in red.
			The plots on the bottom show red stars representing the points around which the refinement is carried out.
		}
		\label{fig:fire}
	\end{figure}
}{\textcolor{red}{Figure kgs not found}}

\IfFileExists{figs_cache_Fire_disorder.pdf}{
	\begin{figure}
		\centering
		\includegraphics[width=0.6\textwidth]{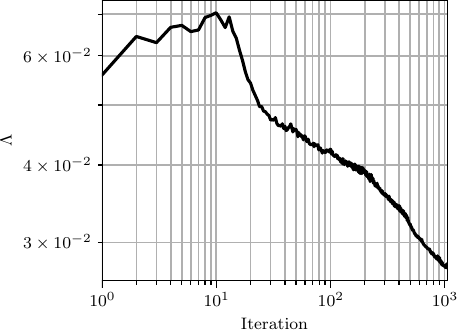}
		\caption{Spatial disorder measure, $\Lambda$, for the particle initialization of a fire emoji with variable resolution.}
		\label{fig:fire-disorder}
	\end{figure}
}{\textcolor{red}{Figure disorder not found}}

\begin{table}
	\centering
	\begin{tabular}{cccccc}
	\toprule
	                        &                & $h_\text{fact}$ & $\max_i{\left(\left|\rho_i - \rho_0\right|\right)}$ & $\max_i{\left(\left|\left|\nabla \Gamma_i\right|\right|_2\right)}$ \\
	Mass Dissipation        & Kernel         &                 &                                                     &                                                                    \\
	\midrule
	\multirow[t]{2}{*}{Yes} & Cubic Spline   & 1.2             & 0.0200                                              & 1.1560                                                             \\
	                        & Quintic Spline & 1.5             & 0.0124                                              & 0.6479                                                             \\
	\cline{1-5}
	\multirow[t]{2}{*}{No}  & Cubic Spline   & 1.2             & 0.0840                                              & 9.1461                                                             \\
	                        & Quintic Spline & 1.5             & 0.1046                                              & 7.1224                                                             \\
	\cline{1-5}
	\bottomrule
\end{tabular}

	\caption{Particle initialization error values for the fire emoji case with variable resolution, with and without mass dissipation.}
	\label{tab:fire-stats}
\end{table}

\IfFileExists{figs_cache_Fire_convsp.pdf}{
	\begin{figure}
		\centering
		\includegraphics[width=0.95\textwidth]{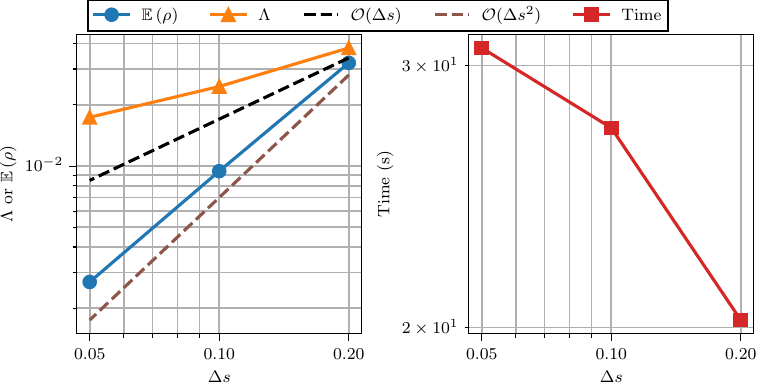}
		\caption{Convergence of the particle initialization of a fire emoji with variable resolution, showing $\Lambda$ and $\mathbb{E}\left(\rho\right)$ on the left and time taken for initialization on the right for different values of $\Delta s$.}
		\label{fig:fire-conv-sp}
	\end{figure}
}{\textcolor{red}{Figure convsp not found}}

\subsection{NACA 0012 Airfoil}
Airfoils are fundamental in many engineering applications involving turbines and aircraft.
In this subsection, we present the results of the particle initialization of a unit chord NACA 0012 airfoil with variable resolution in 2D.
The domain is initially filled with particles at a constant resolution of \(\Delta s = 0.025\).
At the leading and trailing edges, we set \(\Delta s_\text{min} = 0.25 \Delta s\) to target refinement around those regions.
The density and kernel gradient sum are shown in \cref{fig:naca0012}.
The convergence of the spatial disorder measure, \(\Lambda\), is shown in \cref{fig:naca0012-disorder}.
Again, the errors are relatively high compared to the cases with constant resolution.
However, the particle initialization method is able to create distribution that is refined around the leading and trailing edges, and smoothing varying volumes.
Resampling with the quintic spline kernel and \(h_{\text{fact}} = 1.5\) results in significantly lower errors as shown in \cref{tab:naca0012-stats}.

\IfFileExists{figs_cache_Naca_adaptrho.pdf}{
	\IfFileExists{figs_cache_Naca_kgs.pdf}{
		\begin{figure}
			\centering
			\includegraphics[width=0.9\textwidth]{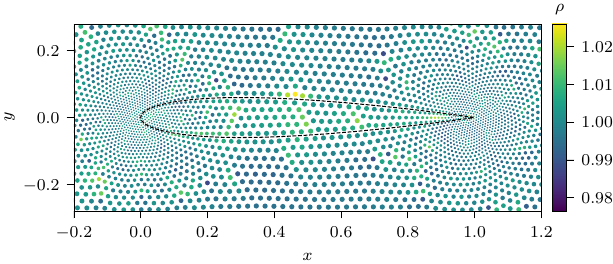}
			\includegraphics[width=0.9\textwidth]{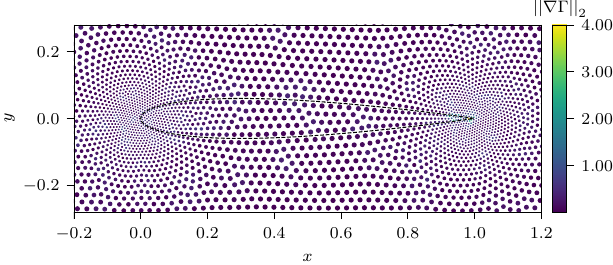}
			\caption{Particle initialization of a NACA 0012 airfoil with variable resolution, coloured by $\rho$~(left) and $\left|\left|\nabla \Gamma\right|\right|_2$~(right).
				The leading and trailing edges are refined.
			}
			\label{fig:naca0012}
		\end{figure}
	}{\textcolor{red}{Figure kgs not found}}
}{\textcolor{red}{Figure adaptrho not found}}

\IfFileExists{figs_cache_Naca_disorder.pdf}{
	\begin{figure}
		\centering
		\includegraphics[width=0.6\textwidth]{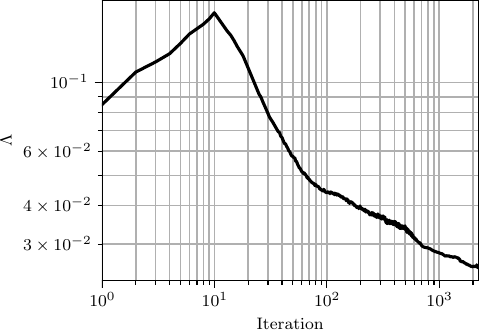}
		\caption{Spatial disorder measure, $\Lambda$, for the particle initialization of a NACA 0012 airfoil with variable resolution.}
		\label{fig:naca0012-disorder}
	\end{figure}
}{\textcolor{red}{Figure disorder not found}}

\begin{table}
	\centering
	\begin{tabular}{ccccc}
	\toprule
	Kernel         & $h_\text{fact}$ & $\max_i{\left(\left|\rho_i - \rho_0\right|\right)}$ & $\max_i{\left(\left|\left|\nabla \Gamma_i\right|\right|_2\right)}$ \\
	\midrule
	Cubic Spline   & 1.2             & 0.0260                                              & 4.0059                                                             \\
	Quintic Spline & 1.5             & 0.0135                                              & 1.3938                                                             \\
	\bottomrule
\end{tabular}

	\caption{Particle initialization error values for a NACA 0012 airfoil with variable resolution.}
	\label{tab:naca0012-stats}
\end{table}

\subsection{Sphere}
\begin{figure}
	\centering
	\includegraphics[width=0.55\textwidth]{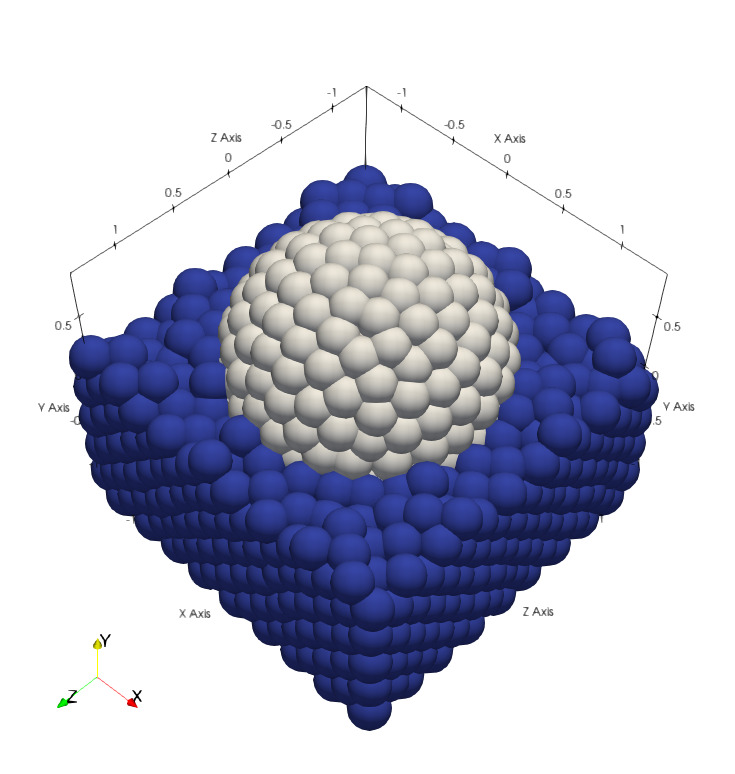}
	\caption{Particle distribution around the packed sphere.
		Some outer particles are clipped to reveal the packed inner sphere.
	}
	\label{fig:sphere-packed}
\end{figure}

A simple sphere is a common test case~\citep{zhuCADcompatibleBodyfittedParticle2021,jiFeatureawareSPHIsotropic2021} used to demonstrate the particle initialization method in 3D.
We use a sphere of unit radius and avoid any refinement.
The particle initialization is performed with a constant resolution of \(\Delta s = 0.2 \).
From \cref{fig:sphere} we can see that the particles have rearranged themselves to form a sphere.
We have shown the sphere along with the outer particles in \cref{fig:sphere-packed}.
Note that we will only show the inner particles to reveal the packed shape in the figures ahead.
The spatial disorder measure, \(\Lambda\), converges to around 0.043 as shown in \cref{fig:sphere-disorder}.
The density variation and the kernel gradient sum are shown in \cref{fig:sphere}.
It is observed that the errors are excellent, just like the non-variable resolution cases in 2D.
The error values are presented with the original Cubic Spline kernel and $h_{\text{fact}} = 1.2$ and the resampled Quintic Spline kernel and $h_{\text{fact}} = 1.5$ in \cref{tab:sphere-stats}.
This case demonstrates that the particle initialization method works well to create a uniformly distributed particle distribution in 3D.

\IfFileExists{figs_cache_Sphere_sphere_rho.pdf}{
	\IfFileExists{figs_cache_Sphere_sphere_kgs.pdf}{
		\begin{figure}
			\centering
			\includegraphics[width=0.47\textwidth]{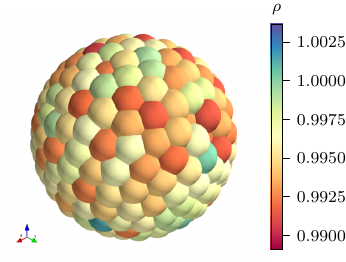}
			\includegraphics[width=0.45\textwidth]{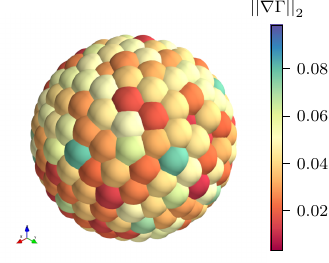}
			\caption{Particle initialization of a sphere with constant resolution, coloured by $\rho$~(left) and $\left|\left|\nabla \Gamma\right|\right|_2$~(right).}
			\label{fig:sphere}
		\end{figure}
	}{\textcolor{red}{Figure kgs not found}}
}{\textcolor{red}{Figure rho not found}}

\IfFileExists{figs_cache_Sphere_sphere_disorder.pdf}{
	\begin{figure}
		\centering
		\includegraphics[width=0.6\textwidth]{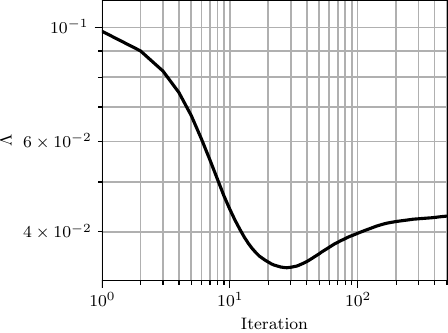}
		\caption{Spatial disorder measure, $\Lambda$, for the particle initialization of a sphere with constant resolution.}
		\label{fig:sphere-disorder}
	\end{figure}
}{\textcolor{red}{Figure disorder not found}}

\begin{table}
	\centering
	\begin{tabular}{ccccc}
	\toprule
	Kernel         & $h_\text{fact}$ & $\max_i{\left(\left|\rho_i - \rho_0\right|\right)}$ & $\max_i{\left(\left|\left|\nabla \Gamma_i\right|\right|_2\right)}$ \\
	\midrule
	Cubic Spline   & 1.2             & 0.0110                                              & 0.0984                                                             \\
	Quintic Spline & 1.5             & 0.0043                                              & 0.0165                                                             \\
	\bottomrule
\end{tabular}

	\caption{Particle initialization error values for a sphere with constant resolution.}
	\label{tab:sphere-stats}
\end{table}

\subsection{Stanford Bunny}
The Stanford Bunny is a widely used 3D test model in computer graphics.
It has served as a standard benchmark for evaluating particle initialization methods~\citep{jiangBlueNoiseSampling2015,negiAlgorithmsUniformParticle2021,zhuCADcompatibleBodyfittedParticle2021}.
The interface particles were generated from an STL file (see \ref{sec:appendix-stl} for details).
A non-varying resolution of \(\left(m/\rho\right)^{1/d} = 0.002\) was used for the particle initialization.
The $\Delta s$ is set to \(0.002\) for all interface particles.
The results of the particle initialization of the Stanford Bunny are shown in \cref{fig:bunny}.
The $\left|\left|\nabla \Gamma\right|\right|_2$ error are shown in \cref{fig:bunny-kgs}.
The spatial disorder measure, \(\Lambda\), converges to around 0.035 as shown in \cref{fig:bunny-disorder}.
The error values are presented with the original Cubic Spline kernel and $h_{\text{fact}} = 1.2$ and the resampled Quintic Spline kernel and $h_{\text{fact}} = 1.5$ in \cref{tab:bunny-stats}.
The details on the bunny's body are well captured, and the particle distribution looks uniform.
Just like the other non-varying resolution cases, the errors and the spatial disorder measure are excellent.
This shows that the particle initialization method is able to create a particle distribution around the complex geometry in 3D.

\IfFileExists{figs_cache_Bunny_bunny_rho.pdf}{
	\begin{figure}
		\centering
		\includegraphics[width=0.95\textwidth]{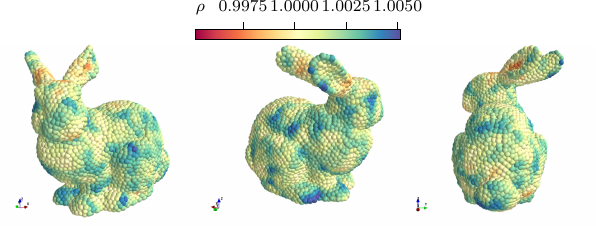}
		\caption{Particle initialization of a Stanford Bunny with constant resolution, coloured by $\rho$.}
		\label{fig:bunny}
	\end{figure}
}{\textcolor{red}{Figure bunny rho not found}}

\IfFileExists{figs_cache_Bunny_bunny_kgs.pdf}{
	\begin{figure}
		\centering
		\includegraphics[width=0.45\textwidth]{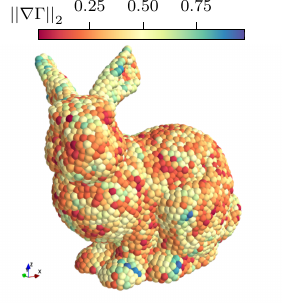}
		\caption{Particle initialization of a Stanford Bunny with constant resolution, coloured by $\left|\left|\nabla \Gamma\right|\right|_2$.}
		\label{fig:bunny-kgs}
	\end{figure}
}{\textcolor{red}{Figure bunny kgs not found}}

\IfFileExists{figs_cache_Bunny_bunny_disorder.pdf}{
	\begin{figure}
		\centering
		\includegraphics[width=0.6\textwidth]{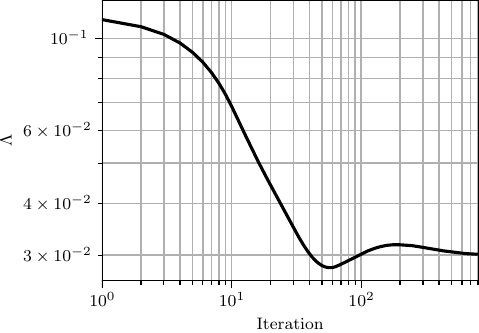}
		\caption{Spatial disorder measure, $\Lambda$, for the particle initialization of a Stanford Bunny with constant resolution.}
		\label{fig:bunny-disorder}
	\end{figure}
}{\textcolor{red}{Figure bunny disorder not found}}

\begin{table}
	\centering
	\begin{tabular}{ccccc}
	\toprule
	Kernel         & $h_\text{fact}$ & $\max_i{\left(\left|\rho_i - \rho_0\right|\right)}$ & $\max_i{\left(\left|\left|\nabla \Gamma_i\right|\right|_2\right)}$ \\
	\midrule
	Cubic Spline   & 1.2             & 0.0154                                              & 9.4745                                                             \\
	Quintic Spline & 1.5             & 0.0062                                              & 1.7641                                                             \\
	\bottomrule
\end{tabular}

	\caption{Particle initialization error values for a Stanford Bunny with constant resolution.}
	\label{tab:bunny-stats}
\end{table}

\subsection{Utah Teapot}
The Utah Teapot, a well-known 3D computer graphics model, is used as a test object for rendering and shading algorithms.
In this work, we use the Utah Teapot to demonstrate the particle initialization method in 3D with variable resolution.
The lid, the handle, and the spout are more intricate and require a finer resolution to capture the details, while the body of the teapot can be represented with a coarser resolution.
The interface particles were generated from an STL file (see \ref{sec:appendix-stl} for details).
The results of the particle initialization of the Utah Teapot with variable resolution are shown in \cref{fig:teapot}.
As intended, the particle initialization method is able to create a particle distribution that is refined around the lid, handle, and spout, maintaining a coarser resolution in the body of the teapot.
The spatial disorder measure, \(\Lambda\) converges to around 0.025 as shown in \cref{fig:teapot-disorder}.
The top x-axis shows the elapsed time in seconds corresponding to the iterations on the bottom x-axis.
Note that this includes the time taken for writing the output files at every 100 iterations along with the time taken for relaxation.
This, however, does not include the time taken for reading the STL file and assigning the properties for the interface particles, which is a one-time cost and insignificant in comparison to the relaxation time.
The error values are presented with the original cubic spline kernel and $h_{\text{fact}} = 1.2$ and resampled quintic spline kernel and $h_{\text{fact}} = 1.5$ in \cref{tab:teapot-stats}.
In summary, this test case demonstrates that the adaptive resolution would be beneficial in places where the corners and crevices have to be resolved with a finer resolution than the rest of the domain, keeping the details of the structure intact while also ensuring that the number of particles is in check.

\begin{table}
	\centering
	\begin{tabular}{ccccc}
	\toprule
	Kernel         & $h_\text{fact}$ & $\max_i{\left(\left|\rho_i - \rho_0\right|\right)}$ & $\max_i{\left(\left|\left|\nabla \Gamma_i\right|\right|_2\right)}$ \\
	\midrule
	Cubic Spline   & 1.2             & 0.0774                                              & 1.2760                                                             \\
	Quintic Spline & 1.5             & 0.0617                                              & 0.2452                                                             \\
	\bottomrule
\end{tabular}

	\caption{Particle initialization error values for a Utah Teapot with variable resolution.}
	\label{tab:teapot-stats}
\end{table}

\IfFileExists{figs_cache_Teapot_teapot_rho.pdf}{
	\IfFileExists{figs_cache_Teapot_teapot_kgs.pdf}{
		\begin{figure}
			\centering
			\includegraphics[width=0.45\textwidth]{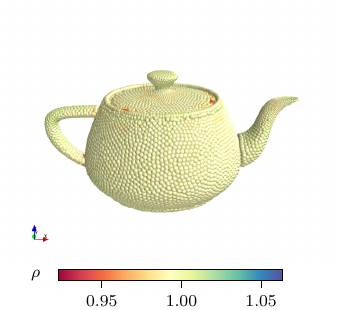}
			\includegraphics[width=0.45\textwidth]{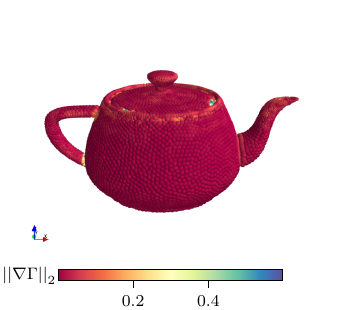}
			\caption{Particle initialization of a Utah Teapot with variable resolution, coloured by $\rho$~(left) and $\left|\left|\nabla \Gamma\right|\right|_2$~(right).}
			\label{fig:teapot}
		\end{figure}
	}{\textcolor{red}{Figure kgs not found}}
}{\textcolor{red}{Figure rho not found}}

\IfFileExists{figs_cache_Teapot_teapot_disorder.pdf}{
	\begin{figure}
		\centering
		\includegraphics[width=0.6\textwidth]{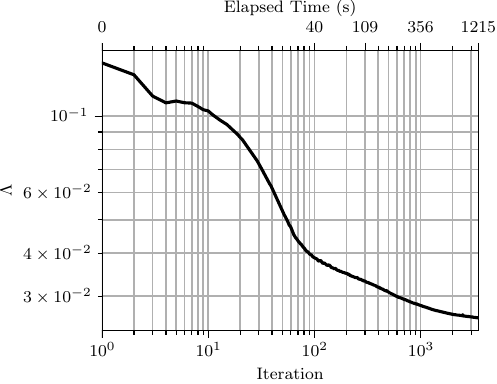}
		\caption{Spatial disorder measure, $\Lambda$, for the particle initialization of a Utah Teapot with variable resolution.
			The top x-axis shows the elapsed time in seconds corresponding to the iterations on the bottom x-axis.
		}
		\label{fig:teapot-disorder}
	\end{figure}
}{\textcolor{red}{Figure teapot disorder not found}}

\subsection{Onera M6 Wing}
The Onera M6 wing is a classic benchmark geometry in aerodynamics, widely used for validating computational methods.
Here, we demonstrate the particle initialization method on the Onera M6 wing with variable resolution in 3D.
The interface particles were generated from an STL file (see \ref{sec:appendix-stl} for details).
The results of the particle initialization of the Onera M6 wing with variable resolution are shown in \cref{fig:onera}.
The spatial disorder measure, \(\Lambda\), converges to around 0.032 as shown in \cref{fig:onera-disorder}.
The details on the leading edge and trailing edge are well captured, the resolution is finer around the leading edge, trailing edge, and the wing tip.
In comparison, the thicker parts of the wing are represented with a coarser resolution.
The variation of $(m/\rho)^{1/d}$ and $h$ is shown in \cref{fig:onera-ds-h}.
$(m/\rho)^{1/d}$ clearly shows the variation of particle spacing around the leading edge, trailing edge, and the wing tip.
The variation of $h$ also shows the same trend.
The error values are presented with the original cubic spline kernel and $h_{\text{fact}} = 1.2$ and the resampled quintic spline kernel and $h_{\text{fact}} = 1.5$ in \cref{tab:onera-stats}.
This shows that the particle initialization method can create a well-distributed particle distribution around the Onera M6 wing, capturing the details of the geometry while maintaining a low density variation and kernel gradient sum.

\IfFileExists{figs_cache_Onera_onera_rho.pdf}{
	\IfFileExists{figs_cache_Onera_onera_kgs.pdf}{
		\begin{figure}
			\centering
			\includegraphics[width=0.45\textwidth]{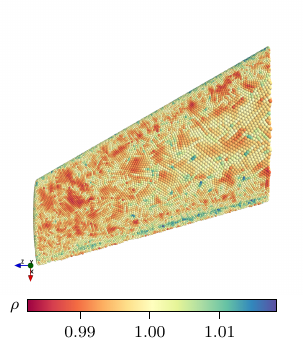}
			\includegraphics[width=0.45\textwidth]{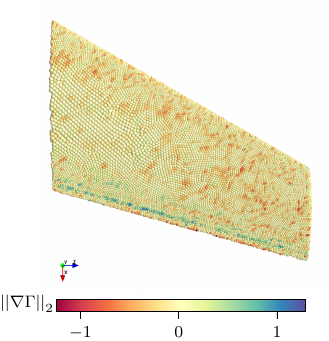}
			\caption{Particle initialization of an Onera M6 wing with variable resolution, coloured by $\rho$~(left), $\left|\left|\nabla \Gamma\right|\right|_2$~(right).}
			\label{fig:onera}
		\end{figure}
	}{\textcolor{red}{Figure kgs not found}}
}{\textcolor{red}{Figure rho not found}}

\IfFileExists{figs_cache_Onera_onera_ds.pdf}{
	\IfFileExists{figs_cache_Onera_onera_h.pdf}{
		\begin{figure}
			\centering
			\includegraphics[width=0.45\textwidth]{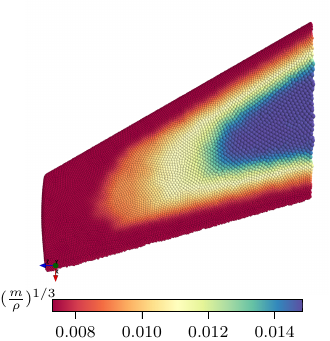}
			\includegraphics[width=0.45\textwidth]{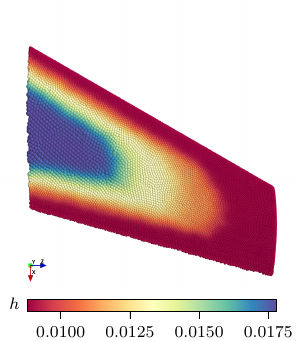}
			\caption{Particle initialization of an Onera M6 wing with variable resolution, coloured by $(m/\rho)^{1/3}$~(left), $h$~(right).}
			\label{fig:onera-ds-h}
		\end{figure}
	}{\textcolor{red}{Figure h not found}}
}{\textcolor{red}{Figure ds not found}}

\IfFileExists{figs_cache_Onera_onera_disorder.pdf}{
	\begin{figure}
		\centering
		\includegraphics[width=0.6\textwidth]{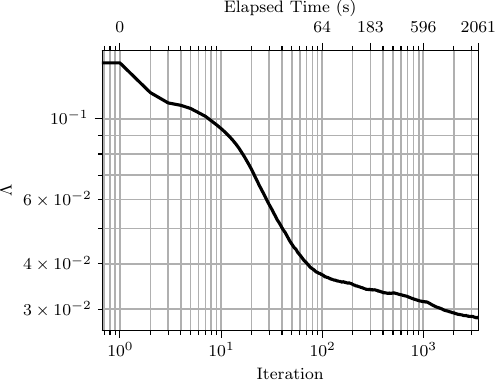}
		\caption{Spatial disorder measure, $\Lambda$, for the particle initialization of an Onera M6 wing with variable resolution.
			The top x-axis shows the elapsed time in seconds corresponding to the iterations on the bottom x-axis.
		}
		\label{fig:onera-disorder}
	\end{figure}
}{\textcolor{red}{Figure ONERA disorder not found}}

\begin{table}
	\centering
	\begin{tabular}{ccccc}
	\toprule
	Kernel         & $h_\text{fact}$ & $\max_i{\left(\left|\rho_i - \rho_0\right|\right)}$ & $\max_i{\left(\left|\left|\nabla \Gamma_i\right|\right|_2\right)}$ \\
	\midrule
	Cubic Spline   & 1.2             & 0.0526                                              & 4.7942                                                             \\
	Quintic Spline & 1.5             & 0.0150                                              & 2.0969                                                             \\
	\bottomrule
\end{tabular}

	\caption{Particle initialization error values for an Onera M6 wing with variable resolution.}
	\label{tab:onera-stats}
\end{table}

\subsection{Ship}
\begin{figure}
	\centering
	\includegraphics[width=0.95\textwidth]{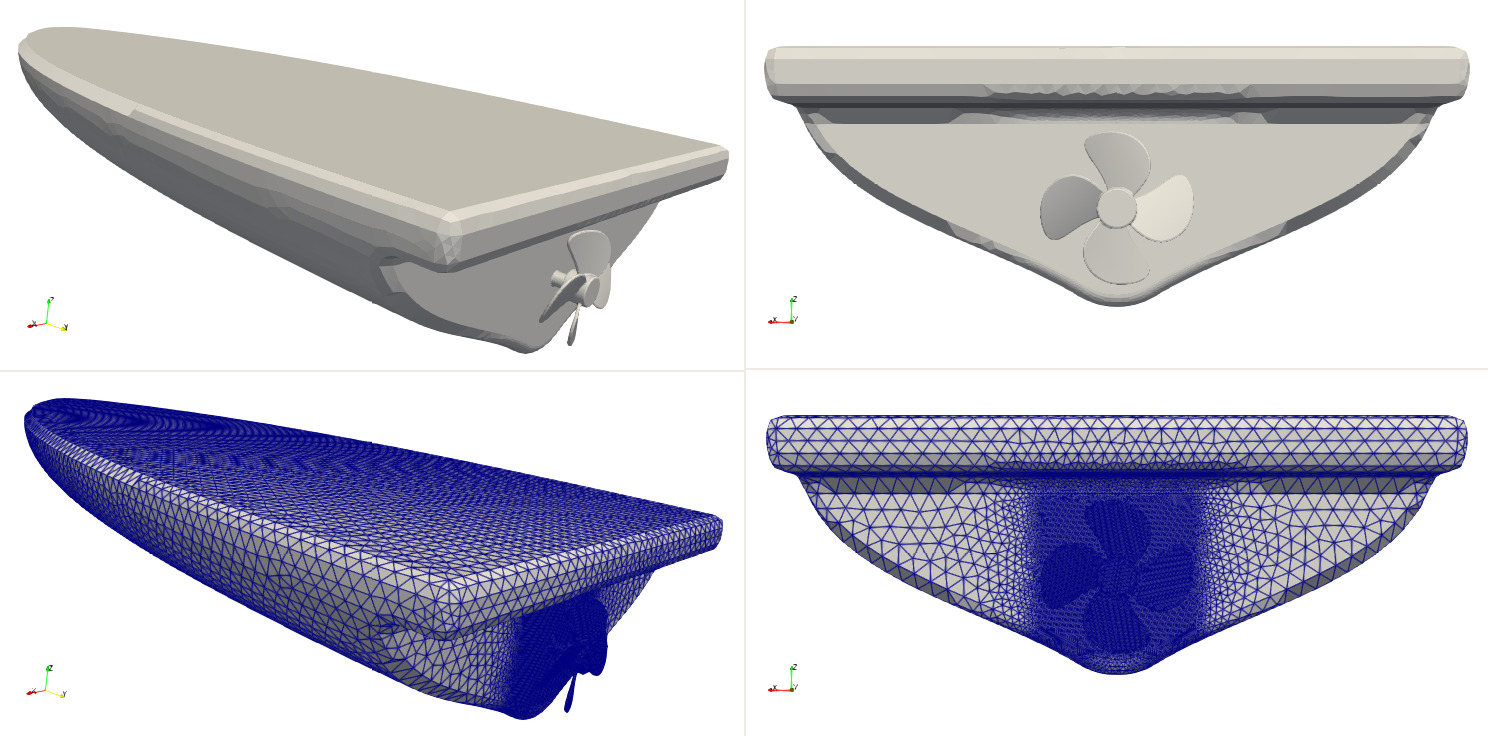}
	\caption{Geometry and the surface mesh of the ship geometry used for particle initialization.}
	\label{fig:ship-stl}
\end{figure}

With this test case, we present the particle initialization around a toy ship geometry with variable resolution in 3D.
This is challenging as the propeller system at the rear of the ship requires fine resolution to capture the details, while the hull has to be represented with a coarse resolution to keep the number of particles in check.
Surface mesh is generated from the created geometry.
The surface mesh is refined near the propeller.
The surface and the surface mesh are shown in \cref{fig:ship-stl}.
The surface mesh is saved as an STL file.
This STL file is used to generate the interface particles and to determine the local resolutions.
The initial set of particles were introduced with $(m/\rho)^{1/3} = 23.56$ and $\Delta s_{\text{min}} = 5.89$.

\begin{figure}
	\centering
	\includegraphics[width=0.3\textwidth]{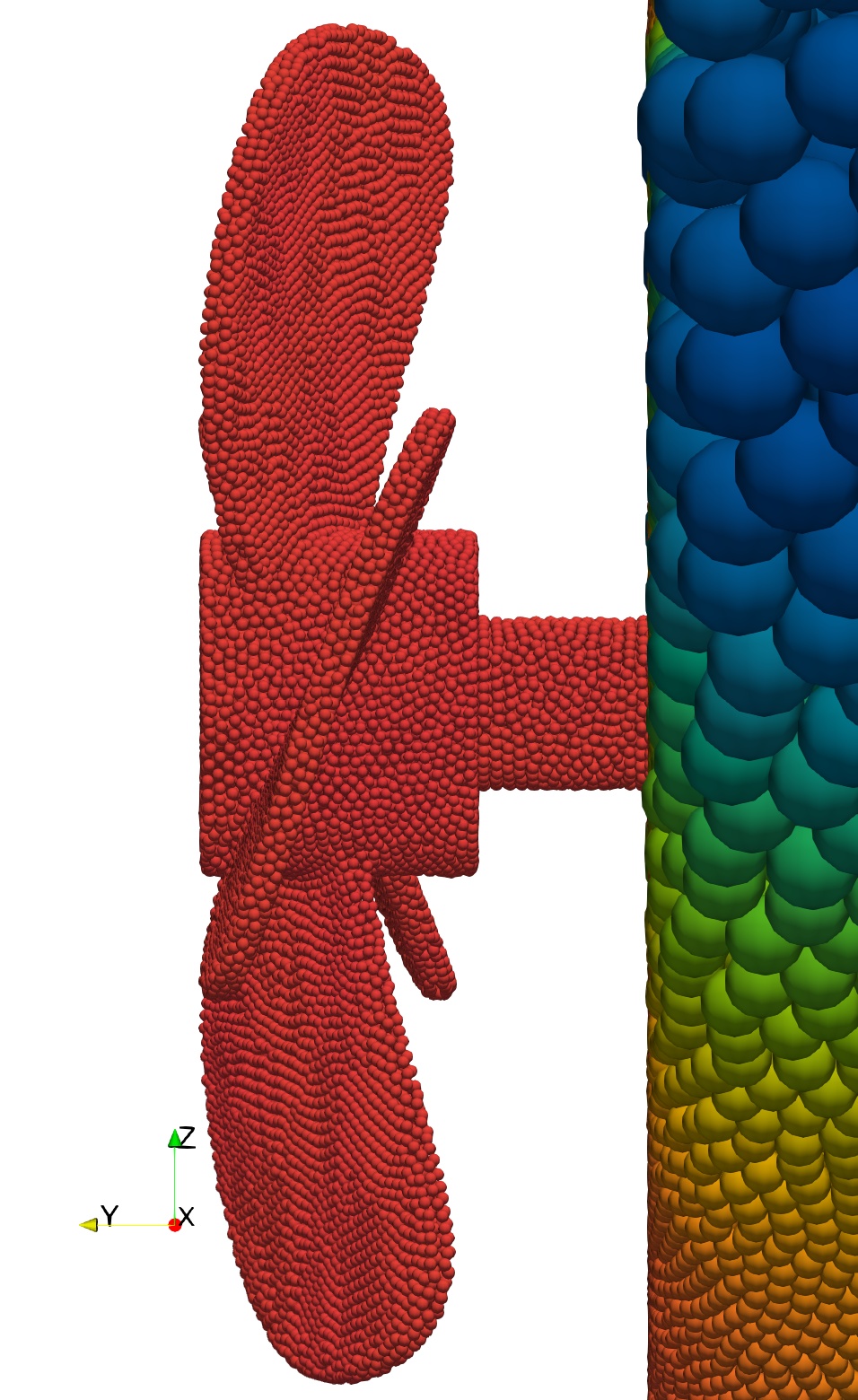}
	\caption{Close-up view of final particle distribution around the ship's propeller.}
	\label{fig:ship-propeller}
\end{figure}

The results of the particle initialization of the ship geometry with variable resolution are shown in \cref{fig:ship}.
The details around the propeller are well captured with a finer resolution, while the hull is represented with a coarser resolution as intended, as can be seen in \cref{fig:ship-propeller}.
The error values are presented with the original cubic spline kernel and $h_{\text{fact}} = 1.2$ and the resampled quintic spline kernel and $h_{\text{fact}} = 1.5$ in \cref{tab:ship-stats}.
The spatial disorder measure, \(\Lambda\), converges to around 0.025 as shown in \cref{fig:ship-disorder}.
The top x-axis shows the elapsed time in seconds corresponding to the iterations on the bottom x-axis.
Note this plot and the similar ones for previous test cases are on a logarithmic scale.
$\Lambda$ and errors drop sharply in the initial iterations.
In \cref{fig:ship-disorder}, it can be observed that $\Lambda$ drops to around 0.03 in about 300 iterations and 474 seconds.
It can also be seen from \cref{fig:ship-evol} that the particles are well-formed at this stage.
So it is fair to say that one can obtain a reasonable distribution in about 300 iterations and 474 seconds.
More iterations reduce $\Lambda$ further, but very slowly.
Considering this, one may argue that it is practical to choose to stop the iterations at this point as further iterations may not be worth the extra computational cost for this case.
However, in general, this decision is application-dependent and is left to be made based on the specific requirements of the simulation.

\IfFileExists{figs_cache_Ship_ship_disorder.pdf}{
	\begin{figure}
		\centering
		\includegraphics[width=0.6\textwidth]{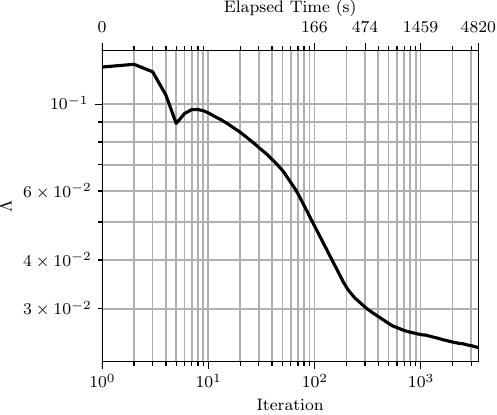}
		\caption{Spatial disorder measure, $\Lambda$, for the particle initialization of a ship geometry with variable resolution.
			The top x-axis shows the elapsed time in seconds corresponding to the iterations on the bottom x-axis.
		}
		\label{fig:ship-disorder}
	\end{figure}
}{\textcolor{red}{Figure ship disorder not found}}

\IfFileExists{figs_cache_Ship_ship_evol.pdf}{
	\begin{figure}
		\centering
		\includegraphics[width=0.95\textwidth]{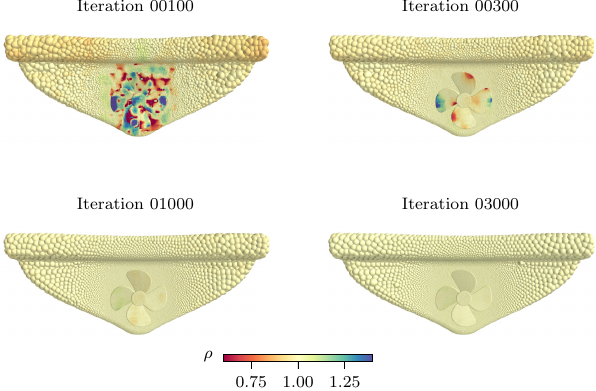}
		\caption{Evolution of the particle initialization of a ship geometry with variable resolution at different iterations, coloured by $\rho$.}
		\label{fig:ship-evol}
	\end{figure}
}{\textcolor{red}{Figure ship evol not found}}

\IfFileExists{figs_cache_Ship_back_rho.pdf}{
	\begin{figure}
		\centering
		\includegraphics[width=0.45\textwidth]{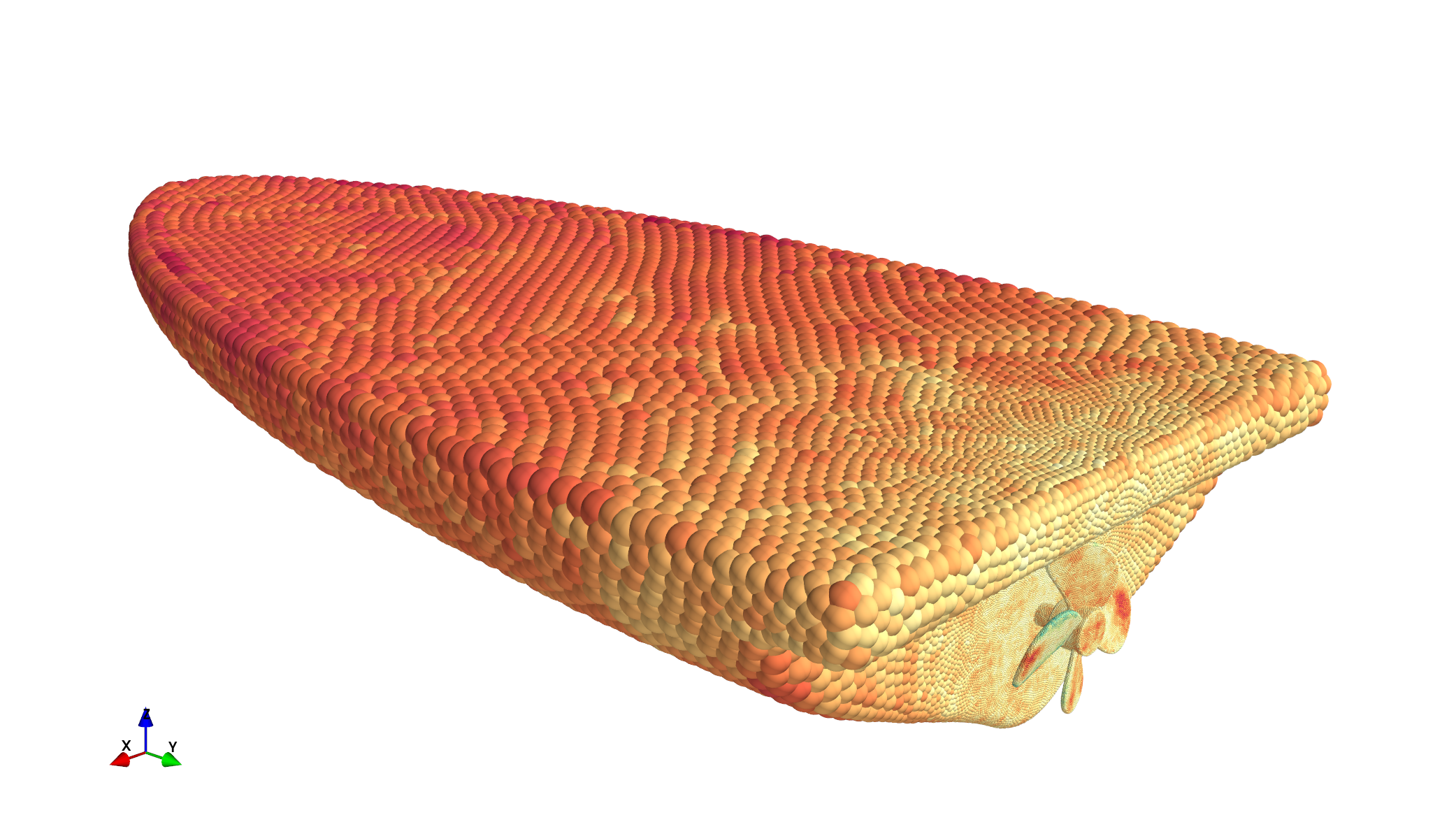}
		\includegraphics[width=0.5\textwidth]{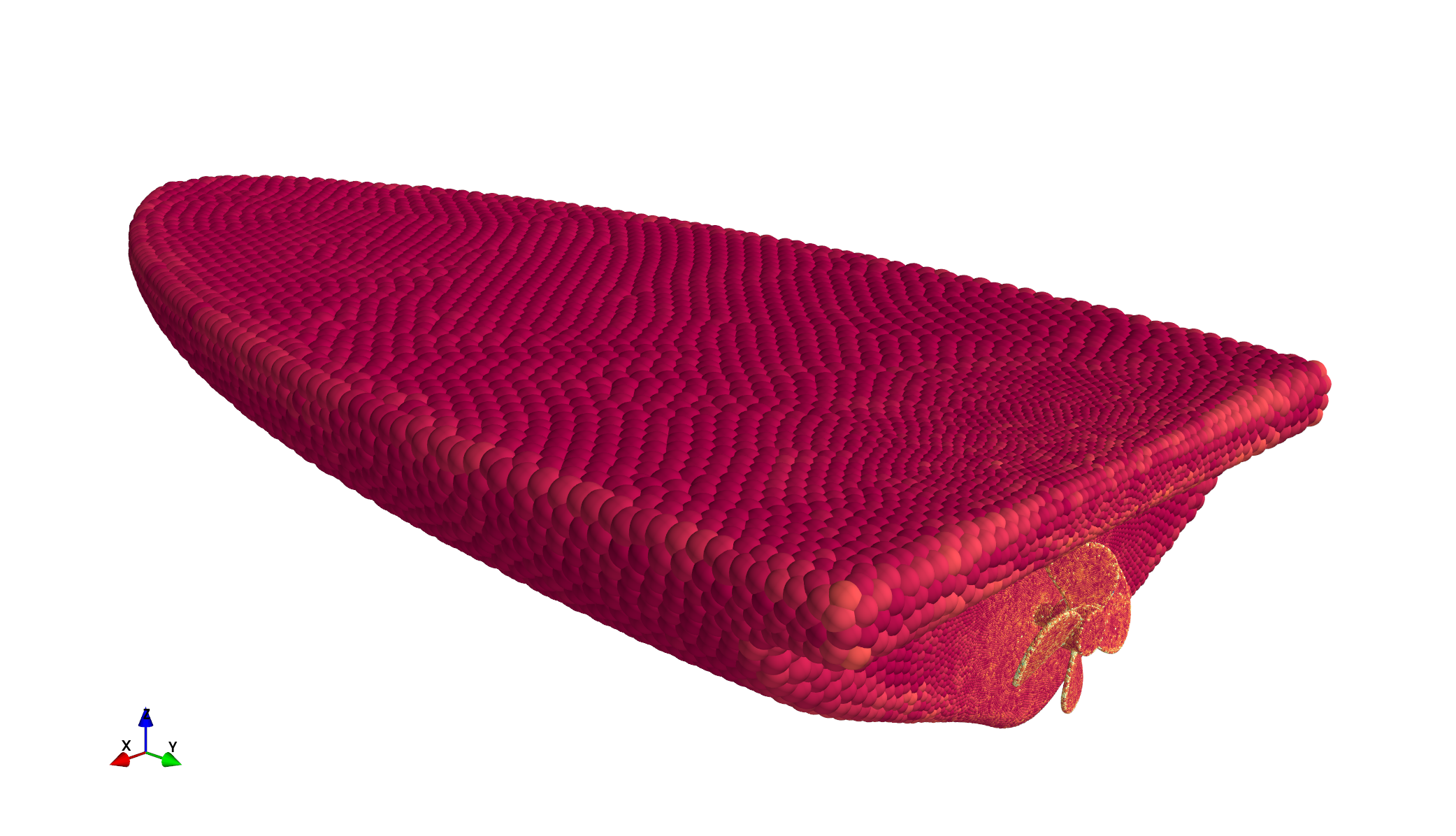}
		\includegraphics[width=0.45\textwidth]{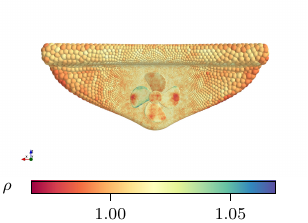}
		\includegraphics[width=0.5\textwidth]{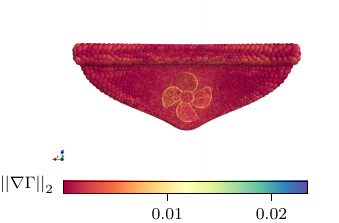}
		\caption{Final result of particle initialization of a toy ship with variable resolution, coloured by $\rho$~(left), $\left|\left|\nabla \Gamma\right|\right|_2$~(right).}
		\label{fig:ship}
	\end{figure}
}{\textcolor{red}{Figure back rho not found}}

\begin{table}
	\centering
	\begin{tabular}{ccccc}
	\toprule
	Kernel         & $h_\text{fact}$ & $\max_i{\left(\left|\rho_i - \rho_0\right|\right)}$ & $\max_i{\left(\left|\left|\nabla \Gamma_i\right|\right|_2\right)}$ \\
	\midrule
	Cubic Spline   & 1.2             & 0.0689                                              & 0.0276                                                             \\
	Quintic Spline & 1.5             & 0.0422                                              & 0.0045                                                             \\
	\bottomrule
\end{tabular}

	\caption{Particle initialization error values for a Ship with variable resolution.}
	\label{tab:ship-stats}
\end{table}

\subsection{Flow over biconvex airfoil for validation} \label{sec:biconvex}
To validate the particle initialization method, we simulate the flow over a unit chord 2D biconvex airfoil at Mach 4.04 and an angle of attack of \(0^\circ\) with $\Delta s = 0.01$.
This is compared with the result of \citet{villodiAdaptiveCompressibleSmoothed2025} in \cref{fig:biconvex-validation}.
The corresponding particle distributions near the leading edge of the airfoil are shown in \cref{fig:biconvex-geom}.
\citet{villodiAdaptiveCompressibleSmoothed2025} used the method of \citet{negiAlgorithmsUniformParticle2021} for particle packing around the airfoil.
We have shown that the results using our particle initialization with a refined resolution of $0.25 \Delta s$ around the leading and trailing edges of the airfoil.
The same flow conditions and particle refinement criteria as \citet{villodiAdaptiveCompressibleSmoothed2025}, i.e., VA-SAS with \(\Delta s_{\min} = \Delta s_{\max} = 10^{-2}\) was used to simulate the flow over the biconvex airfoil.
From \cref{fig:biconvex-validation}, it can be observed that with refinement at the leading edge, the present result avoids the pressure spike at the leading edge.

\IfFileExists{figs_cache_BiconvexAerofoil_biconvex_validation.pdf}{
	\begin{figure}
		\centering
		\includegraphics[width=0.6\textwidth]{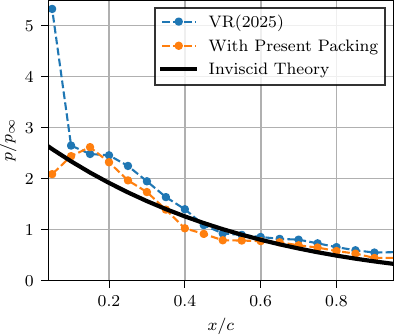}
		\caption{Pressure distribution over a biconvex airfoil at Mach 4.04 and angle of attack of \(0^\circ\).
		VR represents the result of \citet{villodiAdaptiveCompressibleSmoothed2025} case VA-SAS($\Delta s_{\min} = \Delta s_{\max} = 10^{-2}$).
				This result is compared with the result using the present particle packing method.
			}
		\label{fig:biconvex-validation}
	\end{figure}
}{\textcolor{red}{Figure biconvex validation not found}}

\IfFileExists{figs_cache_BiconvexAerofoil_biconvex_geom.pdf}{
	\begin{figure}
		\centering
		\includegraphics[width=0.95\textwidth]{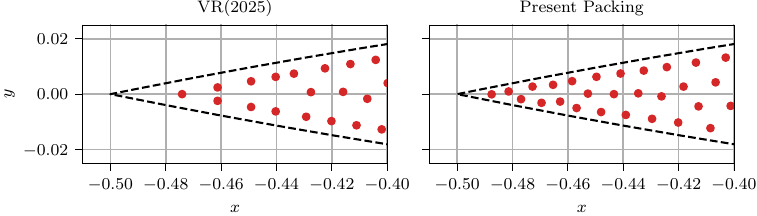}
		\caption{Particle distribution at the leading edge of the biconvex airfoil for the validation case.
		The particles are shown in red and the black dashed line shows the airfoil profile.
		The body particles used in case VA-SAS($\Delta s_{\min} = \Delta s_{\max} = 10^{-2}$) of \citet{villodiAdaptiveCompressibleSmoothed2025} is shown on the left.
				The result of initialization with refinement near the leading edge using the present method is shown on the right.
			}
		\label{fig:biconvex-geom}
	\end{figure}
}{\textcolor{red}{Figure biconvex geom not found}}

\subsection{Discussion on results} \label{sec:discussion}

For the same test cases, the particle initialization method achieves a density variation comparable to the results of \citet{negiAlgorithmsUniformParticle2021} but significantly faster, as shown in Table~\ref{tab:timings}.
The kernel gradient sum, \(\left|\left|\nabla \Gamma\right|\right|_2\) appears to be better than the results of \citet{zhao2025physicsdriven}, though they compute this metric a bit differently.
The errors are different for the test cases in the presented results.
Comparing the presented results with each other, it can be observed that the errors are exaggerated with variable resolution and in 3D.
This is expected as the variable resolution introduces non-uniformity, and 3D increases the degrees of freedom.

Conventionally, kinetic energy has been used as a convergence criterion~\citep{negiAlgorithmsUniformParticle2021, zhuCADcompatibleBodyfittedParticle2021, zhao2025physicsdriven} as the particles in a relaxed configuration tend to have zero velocity.
However, with the present method, the employed mass dissipation results in slow diffusion of mass from higher mass particles in the unrefined regions to the lower mass particles in the refined regions.
This is countered by velocity, i.e, a slow movement of particles in the direction opposite to mass diffusion.
Moreover, in the present scenario, the whole particle movement is not represented by the velocity, as particle shifting is at play.
Therefore, we do not make use of kinetic energy as a criterion in the present method.
The presented results, therefore, are stopped heuristically by carefully observing the evolution of errors.
This also explains the difference in errors and the number of iterations for the different test cases.
However, the simulations using both methods for the timing studies in \cref{tab:timings} are stopped at a precisely defined convergence criterion of 1.5\%.
As shown in \cref{tab:timings}, our approach achieves an order of magnitude speedup compared to the method of \citet{negiAlgorithmsUniformParticle2021}, both in serial and parallel.

In \cref{fig:scaling}, the strong scaling of the present implementation is presented.
The speedup is computed based on the wall time for 100 iterations of the ship case.
Upto 64 threads, each doubling the number of threads yields a speedup of approximately 1.8.
With more than 64 threads, we observe that the other overheads start to dominate.
For parity, the speedups in \cref{fig:scaling} are based on timings on a dual socket node with 48 cores (2 $\times$ Intel Xeon Gold 6240R), the same configuration used for the other 3D examples.
In our trials, we also observed that going from 96 threads on the dual socket node with 48 cores (2 $\times$ Intel Xeon Gold 6240R) to 224 threads on another dual socket node with 112 cores (2 $\times$ Intel Xeon Platinum 8480) yielded a speedup of approximately 2.3 times.

We would like to reiterate the implementation is written using the PySPH framework in Python.
Most of the performance and parallelization aspects are handled by PySPH's internal transpilation pipelines.
Also note that other than compiler's auto-vectorisation capabilities, we are not explicitly targetting Single Instruction Multiple Data~(SIMD) in the present implementation.
We have also not explored distributed memory parallelism or GPU acceleration at this stage.
Therefore, we would like to emphasize that there is room for further performance improvements.

\IfFileExists{figs_cache_ShipScaling_scaling.pdf}{
	\begin{figure}
		\centering
		\includegraphics[width=0.6\textwidth]{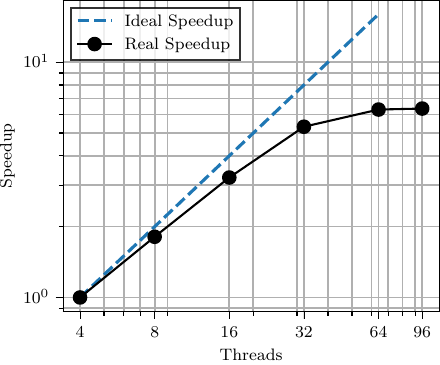}
		\caption{Strong scaling of the present particle initialization method for the ship geometry.}
		\label{fig:scaling}
	\end{figure}
}{\textcolor{red}{Figure scaling not found}}

\begin{table}
	\centering
	\begin{tabular}{lcccccc}
	\toprule
	         & \multicolumn{3}{c}{Serial} & \multicolumn{3}{c}{Parallel (OpenMP)}                                         \\
	Problem  & NR                         & Present                               & Speedup & NR      & Present & Speedup \\
	\midrule
	Circle   & 7.76                       & 0.17                                  & 45.63   & 2.08    & 0.11    & 19.52   \\
	Starfish & 30.39                      & 1.81                                  & 16.79   & 7.68    & 0.96    & 8.03    \\
	Bunny    & -                          & -                                     & -       & 3478.02 & 135.32  & 25.70   \\
	\bottomrule
\end{tabular}

	\caption{Particle initialization timings in seconds for Circle, Starfish, and Stanford Bunny with constant resolution.
		NR stands for \citet{negiAlgorithmsUniformParticle2021}.
		The Stanford Bunny cases were run on a dual socket node with 48 cores~(2 $\times$ Intel Xeon GOLD 6240R).
		For the Circle and Starfish cases, minimum of five runs on an Intel i7-8700 desktop is reported.
	}
	\label{tab:timings}
\end{table}

\section{Summary and concluding remarks} \label{sec:conclusions}

In this work, we have presented a fast and robust method for particle initialization with both constant and variable resolution for complex geometries in 2D and 3D.
The usage of SPH building blocks should make it easy to implement and integrate into existing SPH frameworks.
The approach achieves high-quality, quasi-uniform particle distributions with low spatial disorder and density variation.
However, the method is not without limitations.
Some of these limitations are listed below:
\begin{itemize}
	\item The interface handling procedure assumes that the interface particles are not too close to the frozen particles.
	      The interface being too close to the frozen particles may lead to a deterioration of the quality of the particle distribution locally.

	\item The method assumes a constant density for all particles.
	      This is a limitation for initialization of multi-material domains and variable density bodies or fluids.
	      Though the method does not support variable density inherently, once we obtain the initialized particles, we may scale the masses to achieve the desired density.

	\item The method assumes that the STL files are free of defects.
	      The adaptivity and the relaxation to may work even if the STL geometry is not watertight but the final separation of particles into ``body particles'' and ``surrounding particles'' will run into issues if the holes are large and may require manual intervention.
	      Self-intersection is a more serious issue.
	      We may end up with particles lying very close on the interface with wildly different normals, and/or the normals carried by the interface would just not be initialized correctly.
	      The interface handling part of the algorithm might not work properly in this scenario, producing bad, unusable, or no results.

	\item By defining the regions where refinement is desired, the method automatically varies the resolution from the refined region to the unrefined region.
	      This transition can be controlled using the refinement ratio parameter, \(C_r\).
	      Increasing \(C_r\) will make the transition shorter at the cost of higher errors.
	      The need to maintain this balance eludes exactly targeting desired resolution in nearby regions at will.
\end{itemize}

The presented method uses a global time step as described in \cref{sec:time-integration}.
The refined particles limit the time step for all particles in the domain with their smaller $h$.
Local adaptive time stepping could be used to allow more frequent evolution for the refined particles.
However, merely evolving each particle at its own time step would violate the conservation of mass because of the incompatibility with the mass dissipation equation~(eq.~\ref{eq:mass-exchange}).
This presents a challenge for the implementation of local adaptive time stepping.
Therefore, an area of improvement for future work could be the incorporation of local adaptive time stepping with careful consideration for conservation of mass.

In summary, the proposed method addresses the major pain points of existing particle initialization techniques by providing a simple, efficient, and general solution for the simultaneous initialization of fluid and solid regions with variable resolution.
This paves the way for easy, high-quality particle initialization for meshless simulations.
\section{Acknowledgements}
The authors acknowledge the use of the computing resources of the ACE Facility, Department of Aerospace Engineering, IIT Bombay and Praganak supercomputing facility at IIT Bombay for the computing time.
The authors also acknowledge the National Supercomputing Mission (NSM) for providing computing resources of `PARAM RUDRA' at IIT Bombay, which is implemented by C-DAC and supported by the Ministry of Electronics and Information Technology (MeitY) and Department of Science and Technology (DST), Government of India.
\appendix

\section{Repulsion distance} \label{sec:appendix-repel}
\begin{figure}
	\centering
	\includegraphics[width=0.4\textwidth]{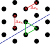} \hspace{0.05\textwidth}%
	\includegraphics[width=0.4\textwidth]{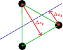}
	\caption{Hexagonal packing in 2D (left) and zoomed view of the triangular repeating unit with distances marked (right).}
	\label{fig:2dhex}
\end{figure}
In this appendix, we show how the distance, \(\Delta s_{d}\) is related to \(\Delta s\).
We assume that the particles are arranged in a hexagonal close packed lattice in 2D.
As shown in \cref{fig:2dhex}, let the distance between two nearest particles be \(\Delta s_n\).
The area of the triangle shown in green is \(\sqrt{3} \Delta s_n^2 / 4\).
The coordination number is 6.
The triangle is formed by three particles with each contributing 1/6 of themselves.
Therefore,
\begin{equation}
	\frac{\sqrt{3}}{4} \Delta s_n^2 = 3 \left(\frac{1}{6}\left(\frac{m}{\rho} \right)\right),
\end{equation}
or
\begin{equation}
	\Delta s_n = \left(\frac{2m}{\sqrt{3}\rho}\right)^{1/2}.
\end{equation}
The distance between close packed planes, denoted as \(\Delta s_p\), is
\begin{equation}
	\Delta s_p = \Delta s_d \cos{\frac{\pi}{6}}  = \left(\frac{2m}{\sqrt{3}\rho}\right)^{1/2} \frac{\sqrt{3}}{2} = \left(\frac{\sqrt{3}m}{2\rho}\right)^{1/2}.
\end{equation}
Substituting \(\Delta s = (m/\rho)^{1/2}\), we get
\begin{equation}
	\Delta s_p = \left(\frac{\sqrt{3}}{2}\right)^{1/2} \Delta s.
\end{equation}
The distance \(\Delta s_d\) is half the distance between two close packed planes.
Therefore,
\begin{equation}
	\Delta s_d = \frac{1}{2} \Delta s_p = \frac{1}{2} \left(\frac{\sqrt{3}}{2}\right)^{1/2} \Delta s = \frac{\sqrt[4]{3}}{2\sqrt{2}} \Delta s.
\end{equation}

Similarly, assuming a cubic close packed lattice in 3D, we arrive at the repulsion distance as
\begin{equation}
	\Delta s_d = \frac{1}{2} \left( \frac{\sqrt[3]{4}}{\sqrt{3}} \Delta s\right) = \frac{\sqrt[3]{4}}{2\sqrt{3}} \Delta s.
\end{equation}

\section{Working with STL files} \label{sec:appendix-stl}

The STL file format is widely used for representing 3D geometries, especially in Computer-Aided Design (CAD) and 3D printing.
It describes the surface geometry of a 3D object using a series of triangular facets.
Each triangle is defined by its three vertices.
STL files can be used to describe the geometry for particle initialization.
We need to create a set of interface particles defining a reference spacing.
For variable resolution, the reference spacing should vary.
The interface particles should also carry normals.

Ideally, the spacing of the interface particles should be similar to the expected spacing of the packed particles.
However, the planar surfaces would be represented with large triangles.
This would result in an undesirably coarse set of interface particles.
In such scenarios, the large triangles can be subdivided by alternatively splitting into three triangles by introducing an additional vertex at the centroid and by splitting into two triangles by introducing an additional vertex at the midpoint of the longest edge.

The regions with intricate details are represented with smaller triangles.
This lends itself as a natural guide for defining the refinement regions for variable resolution particle initialization.
The triangle vertices can be used to define the positions of the interface particles.
The smoothing length, \(h\) for the interface particles can initialised by iterating
\begin{equation}
	\label{eq:stl-h}
	h_i^{n+1} = \frac{1}{2} \left[ \frac{h_i^n}{2} \left( 1 + \sqrt{\frac{N_r}{N_i^n}}\right) + \frac{\sum_{j} h_j^n}{N_i^n} \right].
\end{equation}
This equation is adapted from \citet{yangAdaptiveResolutionMultiphase2019}.
Here, the \(N_i\) is the number of neighbouring interface particles, $N_r$ is the expected number of neighbouring interface particles.
$N_r$ can be set as,
\begin{equation}
	N_r = \lfloor \pi h_\text{fact}^2 \sigma^2 \rfloor
\end{equation}
where \(\sigma\) is the support radius of the kernel.
Once $h$ is initialized, spacing for defining variable resolution initialization can be assigned as $\Delta s_{\min,i} = h_i/h_\text{fact}$.

The interface handling procedure (\cref{sec:interface-handling}) requires normals at the interface particles.
The normals for each triangle can be computed using the cross product of two edges of the triangle.
Each interface particle is then assigned the average of the normals of the triangles to which it belongs.

\bibliographystyle{model6-num-names}
\bibliography{references}

\end{document}